\documentclass[fleqn,usenatbib]{mnras}

\usepackage[T1]{fontenc}

\DeclareRobustCommand{\VAN}[3]{#2}
\let\VANthebibliography\thebibliography
\def\thebibliography{\DeclareRobustCommand{\VAN}[3]{##3}\VANthebibliography}

\usepackage{graphicx}	
\usepackage{amsmath}	
\usepackage{amssymb}	
\usepackage{txfonts}

\title[DM Halo Shape]{Baryons Shaping Dark Matter Haloes}

\author[P. Cataldi et al.]{
P. Cataldi,$^{1}$\thanks{Contact e-mail: pcataldi@iafe.uba.ar}
S. E. Pedrosa,$^{1,2}$
P. B. Tissera$^{3,4}$
and M. C. Artale$^{5}$
\\
$^{1}$Instituto de Astronom\'{\i}a y F\'{\i}sica del Espacio, CONICET-UBA, Casilla de Correos 67, Suc. 28, 1428, Buenos Aires, Argentina\\
$^{2}$Departamento de Física Te\'{o}rica, Facultad de Ciencias, Universidad Aut\'{o}noma de Madrid, E-28049 Cantoblanco, Spain\\
$^{3}$ Instituto de Astrof\'{i}sica, Pontificia Universidad Cat\'olica, Av. Vicuña Mackenna 4860, Santiago, Chile.\\
$^{4}$Centro de Astro-Ingenier\'ia, Pontificia Universidad Cat\'olica de Chile, Av. Vicu\~na Mackenna 4860, Santiago, Chile.\\
$^{5}$ Institut f\"{u}r Astro- und Teilchenphysik, Universit\"{a}t Innsbruck, Technikerstrasse 25/8, 6020 Innsbruck, Austria
}

\date{Accepted 2020 December 22. Received November 30; in original form August 7}

\pubyear{2020}

\begin{document}
\label{firstpage}
\pagerange{\pageref{firstpage}--\pageref{lastpage}}
\maketitle

\begin{abstract}
In this work we aim at investigating the effects of baryons on the dark matter (DM) haloes structure, focusing on the correlation between the presence and importance of stellar discs and the halo shapes. We study the properties of a subsample of DM haloes from Fenix and {\sc eagle} cosmological simulations. We inspect the central regions of haloes in the mass range  $[10.9-992.3] \times 10^{10} \ \rm  M_{\odot}$ at $z=0$, comparing the hydrodynamic runs and their dark matter only (DMo) counterparts. Our results indicate that baryons have a significant impact on the shape of the inner halo, mainly within $\sim$ 20 percent of the virial radius. We find haloes to be more spherical when hosting baryons. While the impact of baryons depends on the mass of the haloes, we also find a trend with morphology which suggests that the way baryons are assembled is also relevant in agreement with previous works. Our findings also indicate that disc galaxies preferentially form in haloes whose DMo counterparts were originally more spherical and with stronger velocity anisotropy. The presence of baryons alter the orbital structure of the DM particles of the haloes, which show a decrease in their velocity anisotropy, towards more tangentially biased  orbits. This relative decrease is weaker in the case of disc-dominated galaxies. Our results point out to a cosmological connection between the final morphology of galaxies and the intrinsic properties of their DM haloes, which gets reinforce by the growth of the discs.
\end{abstract}




\section{Introduction}

\label{sec:intro}

Dark matter (DM) haloes are the building blocks in the concordance cosmological scenario of the Universe. The accurate determination of their properties in the presence of baryons has been a crucial task in the studies of galaxy formation \citep{Knebe2004, Moore,Faltenbacher2005, Avila-Reese, Bailin, Patiri, Vera-Ciro2011}. Additionally, cosmological simulations have been a forceful tool to investigate them. Even though, through the last decade several studies have provided robust insights of the DM haloes characteristics, many problems remain to be fully understood, such as the \textit{"cusp-core"} problem \citep{Moore,Walker2011,Oh2015} or the \textit{"missing satellite problem"}  \citep{Diemand2007,Guo_2010, Bullock-Boylan}.

The outcomes of dark matter only (DMo) simulations showed that CDM haloes are triaxial \citep[e.g.][]{Dubinski,Allgood}. However, the condensation of baryons within the central regions during the assembly of galaxies modifies their internal structure \citep[e.g.][]{Tissera1998,Kazantzidis,Butsky}. 

DM haloes respond by contracting or, in some cases, expanding to the assembly of baryons in the inner regions as shown by comparative studies of the DM profiles in hydrodynamical simulations and those of their DMo counterparts \citep{Pedrosa2010,Tissera2010,Duffy,Dutton,Chua}. Baryonic processes such as supernova feedback (SN) contribute to regulate the presence of baryons in the most inner parts of the galaxy \citep{Navarro1996,Governato}. As a byproduct, SN could also help to transform cuspy into core profiles by triggering cycles of inflows and outflows which can contribute in reshaping the inner regions of haloes \citep{Dutton,Benitez-Llambay}.

The close-connected evolution between the central galaxy and its host halo has been analysed by several studies from
different point of views \citep[e.g.][]{barneswhite1984,Tissera1998,Gnedin,Debattista,Romano2008,Valluri}. In particular, \citet{Pedrosa2010}, \citet{Zavala} and \citet{Teklu} studied the exchange of angular momentum between the galaxy and the inner and outer parts of the DM halo. \citet{Zavala} found a correlation between the evolution of the specific stellar angular momentum of the galaxy and the one of the inner DM halo. \citet{Teklu} reported that disc galaxies are hosted preferentially by haloes that have central angular momentum aligned with the total angular momentum of the halo. More recently \cite{Zhu17} found that when baryons are included, DM particle orbits go mainly from box to tube types. These authors also proposed a link between the angular momentum of the baryonic discs and the DM haloes so that for the changes in orbits to be irreversible, angular momentum exchange between baryons and DM is required \citep[e.g.][]{Pedrosa2010}.

It has been extensively established that baryons also produce an impact on the DM halo shapes. Haloes from hydrodynamical simulations appear to be less triaxials than their DMo counterpart \citep{Tissera1998,Dubinski,Chabrier, Debattista, Kazantzidis, Tissera2010}. The shapes of DM  haloes are related to the orbital structure of the DM particles that form them \citep{Barnes}. The sphericalization detected in the presence of baryons is related with the modification of the DM particles orbits. Using hydrodynamical simulations, \citet{Debattista} and \citet{Valluri} found that box orbits that support triaxial haloes become rounder due to the presence of baryons, causing the shape to change. \cite{Vera-Ciro} analysed the Aquarius simulation haloes with peak circular velocity in the range of $\mathrm{8 \; km/s < V_{max}< 200 \; km/s}$ and found a relation between the triaxiality and the mass of the halo: smallest haloes are between 40 and 50 percent rounder than Milky Way-type ones, at the radius where circular velocity peaks. Additionally \cite{Butsky}, using the NIHAO zoom-in simulations, found a strong mass dependence of the inner halo shape between DMo and Hydro simulations.

\citet{Tissera2010} found that baryons modify the velocity dispersion structure in a rather complicated way, with very different behaviours between haloes of the Aquarius Project \citep{scan2009}. The relation between DM halo shape and velocity anisotropy was also studied by \citet{Sparre2012} for a wide range of DMo simulations finding a correlation between them. It is interesting to point out that the presence of a baryonic disc structure disc-like DM structures. \citet{Schaller2016} showed that for a set of 24 simulated Milky Way like galaxies of the {\sc apostle} and {\sc eagle} projects, 23 of these haloes showed no evidence for a dark disc and the only case found resulted from a recent satellite merger.    

Using the Illustris and Illustris-Dark simulations, \citet{Chua} reported that baryonic physics has a significant impact on the halo shape, sphericalizing haloes, with stronger effects in the inner region. In agreement with \citet{Pedrosa2010} and \citet{Tissera2010}, they found that the presence of baryons altered the DM velocity dispersion and decreased the velocity anisotropy along all radii and masses, their orbits becoming more tangentially biased. 

\citet{Kazantzidis} investigated the effect of the growth of a central disc galaxy on the shape of a triaxial DM halo in a DMo simulation and reported clear sphericalization of the DM haloes by the galaxy discs. Interestingly, \citet{Thob} studied the morphology and kinematics of galaxies in the {\sc eagle} suite, finding a link among several kinematic indicators of baryon morphology. Their results also suggested that there might be an intrinsic correlation between the flatness of the hosted galaxies and their haloes.  

These strong numerical predictions for the three-dimensional shapes of DM haloes (and in particular the Milky Way DM halo) is poorly constrained by current observational data \citep{Bovy2016}. Lately, \citet{Law2010, Bovy2016, Malhan} established stronger observational constrains of the Milky Way halo shape within the galactocentric radial interval (20 kpc < r < 60 kpc), using the effect of Sagittarius stream and the analysis of the observed phase-space of Pal 5 and GD-1, finding that the potential gleaned from the tracks of these streams are precise measurements of the shape of the gravitational potential at the location of the streams, with weaker constraints on the radial and vertical accelerations separately.

In this paper, we extend the analysis done in previous works by studying in detail the connection between galaxy morphology, quantified by the relevance of the disc component, and the properties of the inner regions of DM haloes. We also investigate how this relation changes within the measuring radius and galaxy stellar mass. We explore the possible preexisting correlation between  galaxy morphology and their corresponding DMo haloes. This correlation could be implemented as a "predictor" in semianalytical models of galaxy formation. We use the  disc-to-total stellar mass fraction, $\rm D/T$,  to quantify galaxy morphology with this purpose. The analysis is based on one-to-one comparison between the properties of the DM haloes in the fully hydrodynamical runs {\sc eagle} and Fenix, and their DMo counterparts. Analysing samples from two numerical experiments, which implemented different subgrid physics models, allows us to asses that the trends found do not depend on the particular feedback implementation.

This work is organized as follows.Section~\ref{sec:simu} summarizes the main properties of the simulations. Section~\ref{sec:Results} presents the results of the DM and baryon distribution in the inner regions. The DM shape dependence on galaxy morphology is investigated in Section \ref{sec:shapemorphology}. Finally, we present the conclusions in Section \ref{sec:conclusions}.

\section{Numerical simulations}
\label{sec:simu}

For this study we analyse two simulations performed with different versions of {\sc gadget} code \citep{springel2003,springel2005} and different feedback implementations. In this section we describe their main features.

\subsection{{\sc eagle} simulations}

We analyse galaxies selected from the 100 Mpc sized box reference run of the {\sc eagle} Project, a suite of hydrodynamical simulation that follows the  structure formation in cosmological representative volume. All of them are consistent with the current favoured $\Lambda$-CDM cosmology \citep{Crain2015,Schaye2015}. These simulations include: radiative heating and cooling \citep{Wiersma2009}, stochastic star formation \citep{Schaye2008}, stochastic stellar feedback \citep{Dalla2012} and AGN feedback \citep{Rosas-Guevara2015}. The AGN feedback is particularly important for the evolution of SF activity in massive early-type galaxies (ETGs). An Initial Mass Function (IMF) of \citet{Chabrier} is used. A more detailed description of the code and the simulation can be seen in \citet{Crain2015} and \citet{Schaye2015}.

The adopted cosmological parameters are $\Omega_{m}=0.307$, $\Omega_{\Lambda}=0.693$, $\Omega_{b}=0.04825$, $\mathrm{H_{0}= 100\:  \textit{h} \,  km\,  s^{-1}\,  Mpc^{-1}}$, with $ h=0.6777$ \citep{Planck2013}. The 100 Mpc sized box reference simulation, so called L100N1504, is represented by $1504^{3}$ dark matter particles and the same initial number of gas particles, with an initial mass of $\mathrm{9.70 \times 10^{6}M_{\odot}}$ and $\mathrm{1.81 \times 10^{6}M_{\odot}}$, respectively. The maximum gravitational softening of $0.7 \   \mathrm{kpc}$ is adopted. 

The energy injected into the gas corresponds to $ 10^{51}$ erg per supernovae event times a dimensionless factor, $f_{E}$, that depends on the local gas metallicity and density. For haloes with more than 1500 DM particles (approximately $\mathrm{ M}>10^{10}h^{-1} \rm M_{ \odot }$), black hole sink particles are placed at the center of the halos. Feedback due to AGN activity is implemented in a similar way to the feedback from star formation. For haloes, at $\rm z=0$,  with masses higher than $\mathrm{>10^{13}M_{\odot}}$ the stellar feedback becomes less effective but AGN feedback can still expel baryons \citep{SchallerFeedback}.

The halo catalogue \footnote{We use the publicly available data by \citet{McAlpine} http://icc.dur.ac.uk/Eagle/database.php} was constructed by using a \rm{Friends-of-Friends} algorithm, while the substructures were identified using the SUBFIND algorithm \citep{springel2001}. For the {\sc eagle} simulations the match of the haloes in the Hydro and DMo simulations was done following \citet{Schaller2015}.

\subsection{Fenix simulations}

We use the cosmological simulation S230D from the Fenix Project suite. Several properties of galaxies in this simulation have been thoroughly studied such as the morphological properties \citep{Pedrosa2014}, the size-mass relation and the angular momentum evolution \citep{Pedrosa2015}, the stellar and gaseous metallicity gradients of the disc components \citep{Tissera2016a,Tissera2016b,Tissera2017}, the chemical abundance of the circumgalactic medium \citep{Machado2017} and the fundamental properties of the elliptical galaxies, such as  Faber-Jackson relations and the Fundamental Plane \citep{Rosito}.

This simulation is consistent with a $\Lambda$-CDM universe with $\mathrm \Omega_{\mathrm m}=0.3$, $\mathrm \Omega_{\mathrm \Lambda}=0.7$ and $\mathrm \Omega_{\mathrm b}=0.04$, $\mathrm H_{0}=100 \ h \mathrm{ \ km \ s^{-1}  \ Mpc^{-1}} $ with $h=0.7$, and a normalization of the power spectrum of $\mathrm \sigma_{8} = 0.9$. The simulated box is $14 \mathrm{\ Mpc}$ a side. The initial conditions have  $2 \times 230^{3}$ total particles and a mass resolution of  $4.3 \ \times{}10^{6} \mathrm M_{\odot}$ and $6.4 \times{}10^{5} \mathrm M_{\odot}$ for the DM particle and initial gas particle, respectively. The maximum gravitational softening is $0.35 \  \mathrm{kpc} $. The initial condition has been chosen to correspond to a typical region of the Universe with no massive group present (the largest haloes have virial masses smaller than $\sim \mathrm{10^{13}M_{\odot}}$). We acknowledge the fact that the initial conditions (ICs) represent a small volume of the Universe. Nevertheless, \citet{DeRossi2013} showed that the growth of the simulated haloes are well-described in these simulations by confronting them with those from the Millenium Simulation  \citep{Fakhouri}.

These simulations were run using {\sc gadget-3}, an update version of {\sc gadget-2} \citep{springel2003, springel2005}, optimised for massive parallel simulations of highly inhomogeneous systems. It includes treatments for metal-dependent radiative cooling, stochastic star formation, chemical and energetic supernovae (SN) feedback \citep{scan2005, scan2006}. This feedback model is able to reproduce galactic mass-loaded winds without introducing any mass-dependent scaling parameter. It also includes a multiphase model for the ISM that allows the coexistence of the hot, diffuse phase and the cold, dense gas phase \citep{scan2006, scan2008}, where star formation takes place. Part of the stars ends their lives as Type II and Type Ia Supernovae, injecting energy and chemical elements into the ISM. Each SN event releases $7\ \times \ 10^{50}$ erg, which are distributed equally between the cold and hot phases surrounding the stellar progenitor. The adopted code uses  the chemical evolution model developed by \citet{mosco2001} and adapted to {\sc gadget-3}  by \citet{scan2005}. This model considers the enrichment by SNII and SNIa adopting the yield prescriptions of \citet{WW95}  and \citet{iwamoto1999}, respectively and the Initial Mass Function (IMF)  of \citet{Salpeter}. The synthesised chemical elements are distributed  between the cold and hot phase ($\mathrm{80 \%}$ and $\mathrm{20 \%}$, respectively). The lifetimes for SNIa are randomly selected within the range [0.1, 1] Gyr. Albeit simple, this model reproduces well the mean chemical patterns obtained by adopting the single degenerated model \citep{Jimenez2015}. This combination of star formation and feedback parameters produced systems which can reproduce the angular momentum and the mass-size relation of the discs and bulges of galaxies in agreement with the observational trends \citep{Pedrosa2015}.

The halo catalog was constructed by using a \rm{Friends-of-Friends} algorithm, while the substructures were identified using the {\small SUBFIND} algorithm \citep{springel2001}. As a result 317 galaxies were identified, resolved with more than $2000$ DM particles. The virial masses of the haloes ($\rm M_{200}$) were defined as the mass within a sphere of radius ($\rm R_{200}$) containing$\sim 200$ times the cosmic critical matter density at the corresponding redshift. The hydrodynamic simulation has a DMo run counterpart. We refer to them as "Hydro" and "DMo", respectively. The Hydro and DMo simulations start from identical ICs. The shrinking sphere method proposed by \citet{Power2003} is applied to find the coordinates of the centre of mass of the haloes in each simulation. Haloes are matched between the Hydro and DMo runs by one-by-one as performed for the EAGLE set.

\section{Properties of DM Haloes}
\label{sec:Results}

In order to minimize numerical artifacts, we select objects resolved with more than $10000$ baryonic particles within the optical radius ($\rm R_{\rm opt}$\footnote{The optical radius, $\rm R_{\rm opt}$, is defined as the radius that encloses 80 percent of the baryonic mass (gas and stars) of the galaxy.}). This threshold yields a subsample of 38 objects for the Fenix Project simulation, with masses in the range $[10.9-254.6] \times 10^{10}  \ \rm M_{\odot}$. While for {\sc eagle} simulation, the subsample comprises 1696 objects from the original sample selected by \citet{Tissera2019}, with masses in the range  $[15.1-992.3] \times 10^{10} \ \rm  M_{\odot}$. The DM haloes were re-scaled in order to take into account the baryonic fraction  $1-\rm f_{bar}$, where $\rm f_{bar}=\Omega _{b}/\Omega _{m}$, when comparing Hydro and DMo runs.

The Fenix haloes have radial density profiles that are well described by an \citet{Einasto} and Navarro-Frenk-White (NFW) \citep{Navarro1996,Navarro2} profiles for Hydro and DMo runs respectively. In the case of the {\sc eagle} sample, \citet{Schaye2015} find that they are well-fitted by a sum of a NFW profile in the outer regions and an NFW-like profile but with a sharper bend, that takes into account the effect of the accumulation of stars in the inner regions.

\subsection{Baryons distribution in the inner regions}
\label{sec:propieties}

The modification of the potential well due to the gathering of baryons in the inner regions to form a galaxy, induces changes in the mass distribution of the DM haloes. 
Several studies have addressed this issue adopting different radius to evaluate the  effects \citep{Butsky,Zavala,Thob,Chua}. In order to estimate the radius that maximizes the signal, we inspect three possible selections defined by the radii that enclose 5, 10 and 20 percent of M$_{200}$ (hereafter measuring radius).

\begin{figure}
   \centering
     \includegraphics[width=0.7\columnwidth]{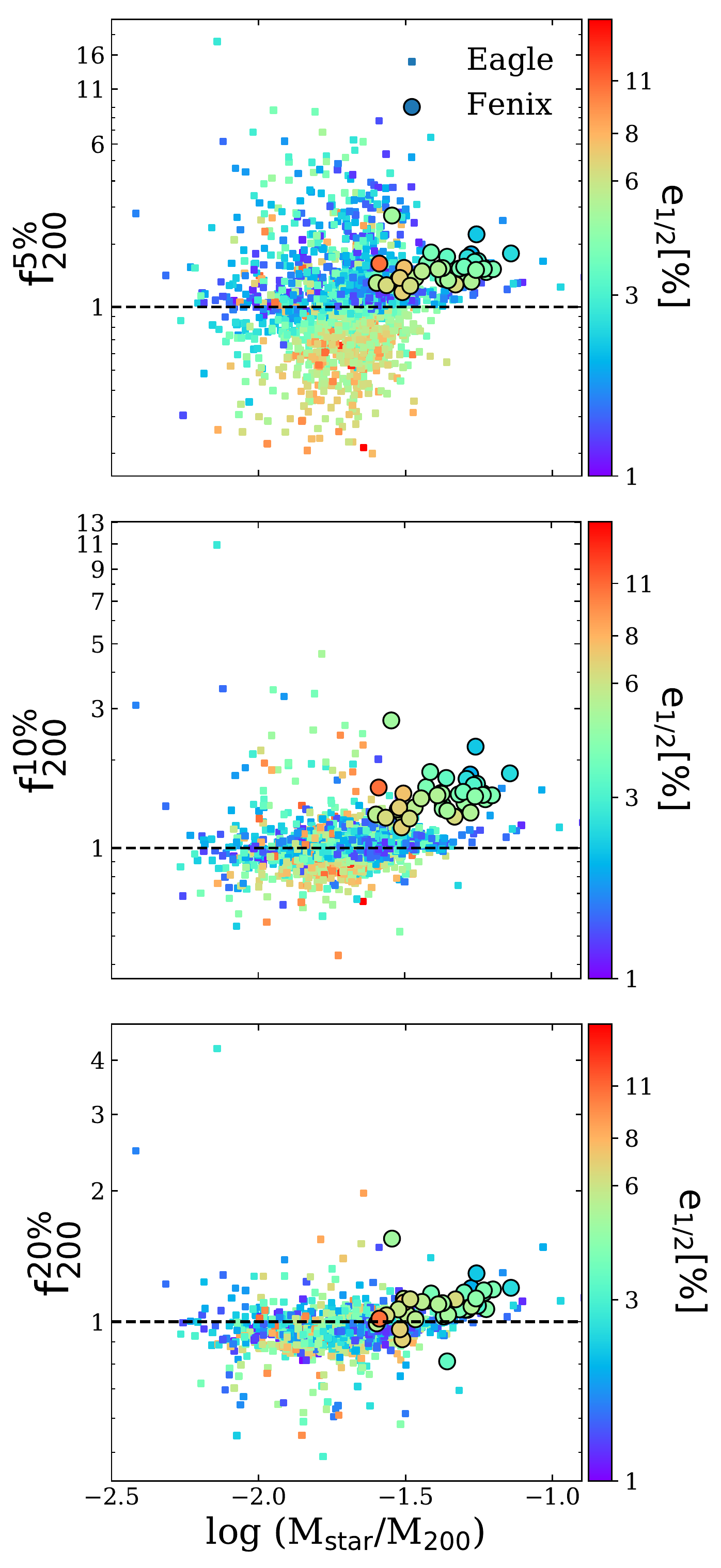}
     \caption{Halo DM mass ratio  ( $\rm f_{_{200}}^{^{ i\%}}$= \rm M$^{\rm Hydro}$/M$^{\rm DMo}(\mathrm{i\% R_{200}})$) as a function of the ratio stellar-to-halo mass $\rm log(M_{ \rm star}/M_{ \rm 200})$. Points are colour coded by galaxy compactness percentage $\rm e_{\rm 1/2}\left [ \% \right ]$ at 5, 10, and 20 percent of R$_{200}$ (top, middle, and lower panel, respectively). We find a trend for less extended galaxies to be more contracted in the inner radius.}
   \label{fig:fig1}
 \end{figure}

This redistribution can be appreciated in Fig. \ref{fig:fig1} where we show the ratio between the DM masses measured at the three defined measuring radii in the Hydro and DMo runs, $\rm f_{_{200}}^{^{ i\%}}$= \rm M$^{\rm Hydro}$/M$^{\rm DMo}(\mathrm{i\% R_{200}})$, as a function of the stellar-to-halo mass ratio $\rm log(M_{ \rm star}/M_{ \rm 200})$. The label $i$ denotes the measuring radii defined to enclose 5, 10, and 20 percent of M$_{200}$.

The results in Fig. \ref{fig:fig1} show that, for the inner regions, haloes in {\sc eagle} can contract or expand (i.e. higher or lower DM masses in the Hydro run compared to the DM only). This type of behaviour has also been detected by \citet{Dutton}  using 100 hydrodynamical cosmological zoom-in simulations of the NIHAO Project. While the contraction is the response to the increase of the potential well due to the accumulation of baryons, the expansion is claimed to be the result of stellar feedback.  

An important fraction of the {\sc eagle} haloes expand when comparing Hydro and DMo runs, while Fenix haloes contract in almost all cases and radii. The expansion in {\sc eagle} haloes in the central regions has been thoroughly  analysed by \citet{SchallerFeedback} and is the result of a more effective stellar feedback and a greater loss of baryons \citep{Sawala2013}. Also it can be noticed in Fig. \ref{fig:fig1} that for higher masses, the contraction is more important as a result of a less effective feedback action. As can be seen, in the case of Fenix simulation, the subsample of haloes studied always contracts when baryons are present within the mass range obtained with this experiment.

We also estimate  the galaxy compactness factor,  $\rm e_{\rm 1/2}\left [ \% \right ]$ \citep{Dutton}, defined as the ratio between the galactic half-mass radius\footnote{The  half-mass radius, $\rm r_{\rm 1/2}$, is defined as the one that enclosed 50 percent of the baryons.}, $\rm r_{\rm 1/2}$,  and the virial radius, i.e. $\mathrm{(100*r_{1/2})/R_{200}}$. For convenience it is expressed as a percentage. The colours in Fig. \ref{fig:fig1} map $\rm e_{\rm 1/2}\left [ \% \right ]$. In the most inner regions, within 5 percent of $\mathrm{R_{200}}$, haloes hosting more compact galaxies tend to be more concentrated as can be seen in the top panel of Fig. \ref{fig:fig1}. When these ratios are measured at $\mathrm{0.10R_{200}}$  ($ \rm f_{_{200}}^{^{10 \%}}$) and $\mathrm{0.20R_{200}}$ ($ \rm f_{_{200}}^{^{20 \%}}$), this trend gradually disappears. This is expected since the larger effects on baryons in the density distribution will be in the very central region where they are located. The concentration of the most inner regions of the haloes depends not on the global baryonic mass gathered in the very centre of haloes but also on their degree of compactness \citep[e.g.][]{Pedrosa2010,Tissera2010,Dutton}.

The change of the DM concentration in the inner regions can be also quantified through the contraction/expansion parameter $\mathrm{\Delta_{\rm v/2}}$, defined as the ratio of the mean DM density to the critical density $\rho_{crit}$ within the radius at which the DM reaches half of its maximum circular rotational velocity, $\mathrm{R _{\rm V/2}}$.

This parameter was  previously implemented in \citet{Pedrosa2010} in the study of the halo contraction evolution with the redshift and originally introduced by \citet{Allam2002}. By using the radius "half way up", $\mathrm{R _{\rm V/2}}$, the rising part of the rotation curve, \citet{Allam2002} focus on the region where conflicts between predicted and observed DM halo densities are more severe (and where the observations are still typically robust against resolution uncertainties).

\begin{figure}
 \centering
 \includegraphics[width=0.67\columnwidth]{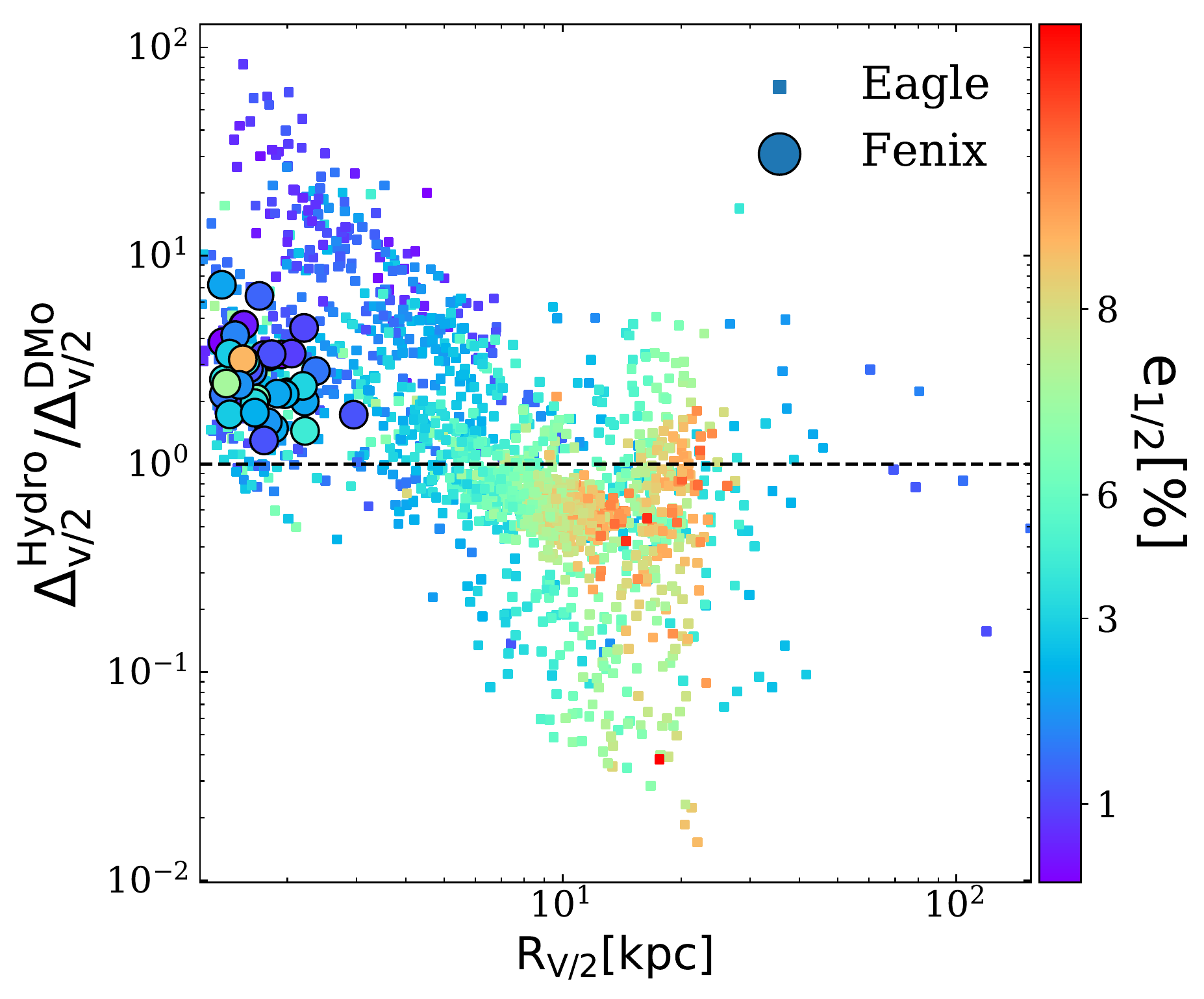}
 \includegraphics[width=0.67\columnwidth]{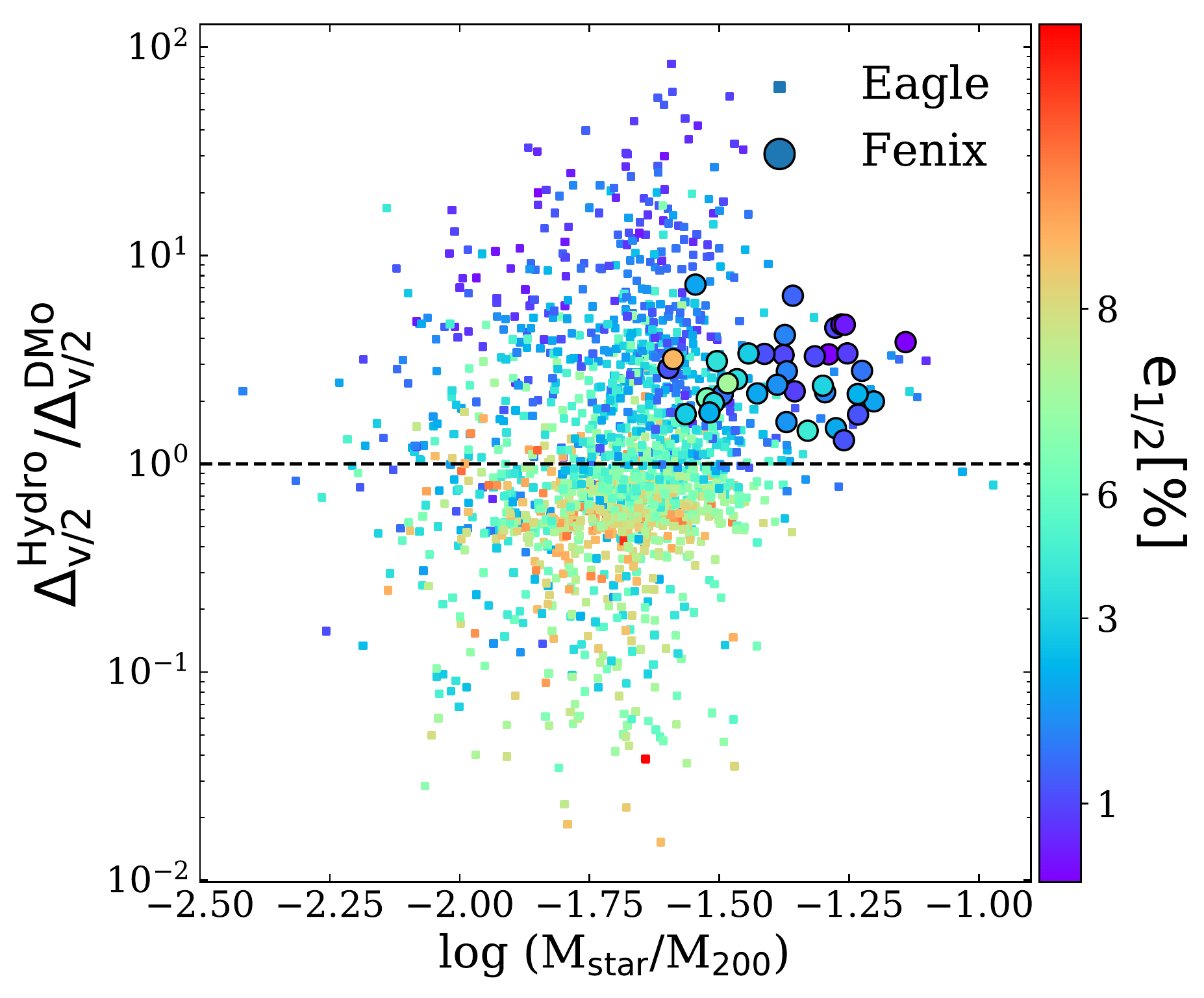}
 \caption{Halo mass contraction/expansion, $\Delta _{\rm v/2}^{\rm Hydro}/\Delta _{\rm v/2}^{\rm DMo}$, as a function of $\rm R_{V/2}$ radii (upper panel) and the  stellar-to-halo-mass ratio $\rm log(M_{ \rm star}/M_{ \rm 200})$ (lower panel). Colours map the galaxy compactness percentage $\rm e_{\rm 1/2}\left [ \% \right ]$ .}
 \label{fig:fig2}
\end{figure}

In Fig. \ref{fig:fig2} we compare the concentration parameters from the Hydro and DMo runs as a function of $\mathrm{R _{\rm V/2}}$ and the stellar-to-virial-mass ratio. Haloes with higher contraction (i.e., $\Delta _{\rm v/2}^{\rm Hydro}/\Delta _{\rm v/2}^{\rm DMo} \gg 1$) are populated with more compact galaxies (i.e., lower $\rm e_{\rm 1/2}\left [ \% \right ]$). The compactness of galaxies appears as a key parameter to trace the degree of concentration of the dark matter haloes. When haloes expand (i.e., $\Delta _{v/2}^{\rm Hydro}/\Delta _{v/2}^{\rm DMo} \ll 1$), they tend to host more extended galaxies. In the top panel it can be seen that compact galaxies are hosted by haloes with shorter $\mathrm{R _{\rm V/2}}$, indicating that the DM maximum rotational velocity $V_{\rm max}$ moves to the inner part of the halo with respect to their DMo counterparts. The lower panel shows that both Fenix and {\sc eagle} subsamples present a slight correlation (Spearman rank coefficient $\rho=0.15$) with $\rm log(M_{ \rm star}/M_{ \rm 200})$, in the sense that more massive galaxies  with respect to halo mass tend to gather also larger DM masses within the very inner regions when baryons are present.

These results confirm the discussed above findings, showing that less concentrated galaxies are located in haloes with expanded central regions with respect to their DMo counterparts. This is clear for the {\sc eagle} halo sample, which in turn spans on a wide range of stellar-to-dark-matter ratio.

\subsection{Baryons shaping haloes}
\label{sec:shape}

Our main interest is to dig into the interdependence of halo shape and galaxy morphology  and try to quantitatively correlate them.

We describe the shapes using the  semi-axes of the triaxial ellipsoids, $\rm a>b>c$, where $\rm a$, $\rm b$ and $\rm c$ are the major, intermediate and minor axis respectively of the $shape$ $tensor$ $\rm S_{ij}$ \citep[e.g.][]{Bailin, Zemp}. An iterative method is used, starting with particles selected in a spherical shell \citep[i.e. $ \rm q=s=1$][]{Dubinski, Curir1993}.  

In order to obtain these ratios $\rm q\equiv b/a$ and $\rm s\equiv c/a$,  we diagonalize the reduce inertia tensor to compute the eingervectors  and eigenvalues, as described in  \citet{Tissera1998}. Traditionally the $\rm s$ shape parameter has been used as a measure of halo sphericity \citep[e.g.][]{Allgood,Vera-Ciro, Chua}.

We adopt the triaxiality parameter, defined as $\rm T \equiv (1-q^{2})/(1-s^{2})$, which quantifies the degree of prolatness or oblatness: $\rm T=1$ describes a completely prolate halo ($\rm a > b\approx c$) while $\rm T = 0$ describes a completely oblate halo ($\rm a\approx  \rm b > c$). Haloes with $\rm T > 0.67$ are considered prolate and haloes with $\rm T < 0.33$ oblates, while those with $\rm 0.33 <T <0.67$ are considered triaxials \citep{Allgood,Artale2018}.  

In the next sections we inspect the shape parameters of the haloes in relation with the measuring radius of the inner region, the D/T fraction and the stellar mass of the hosted galaxy. 

\subsubsection{Shape dependence on radius} \label{ShapeRadio}

To characterize galaxy morphology we use the $\rm D/T$ ratio, calculated for the {\sc eagle} galaxies in \citet{Rosito}. These authors adopted the criteria of \citet{Tissera2012}. This implementation used the circularity parameter defined as $\mathrm{\epsilon=J_{z}/J_{z,max}(E)}$, the ratio between the angular momentum of each stellar particle and $ \mathrm{J_{z,max}(E)}$, the maximum $\mathrm{J_{z}}$ over all particles at a given binding energy, E. A star on a prograde circular orbit in the disc plane has  $\epsilon \simeq 1$. The disc component is associated with those particles with $\epsilon > 0.5$ and the rest of the particles are associated with the spheroidal component. The same procedure is applied to Fenix galaxies. This implementation allows us to define the morphology of a galaxy by using the stellar disc-to-total mass ratio, D/T.

\begin{figure*}
\centering
  \includegraphics[width=\textwidth]{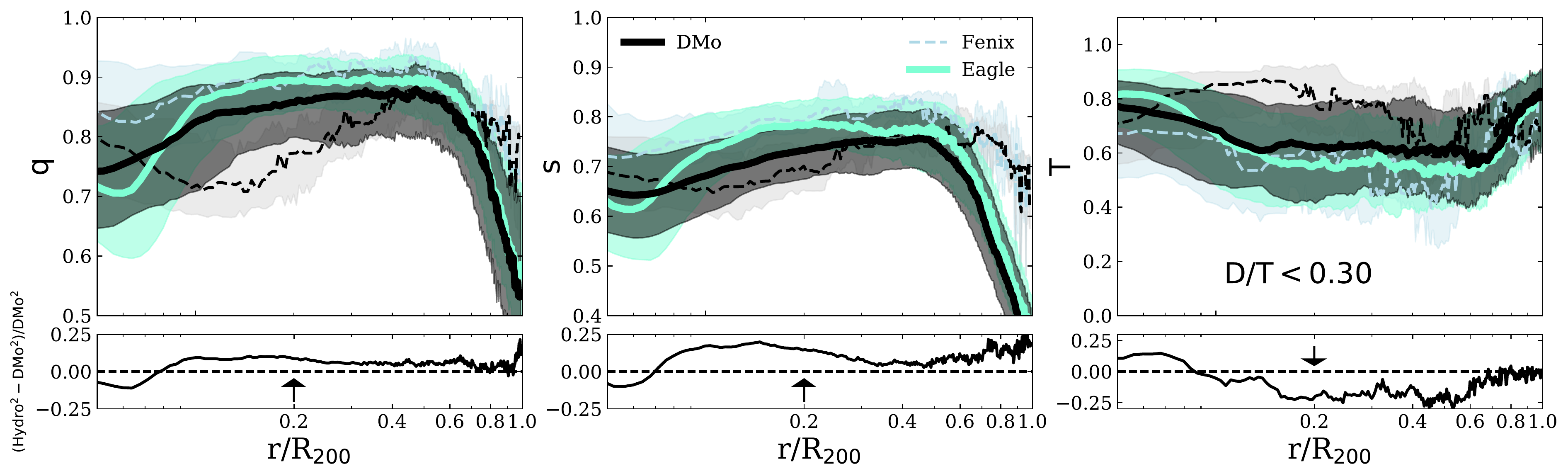}
  \includegraphics[width=\textwidth]{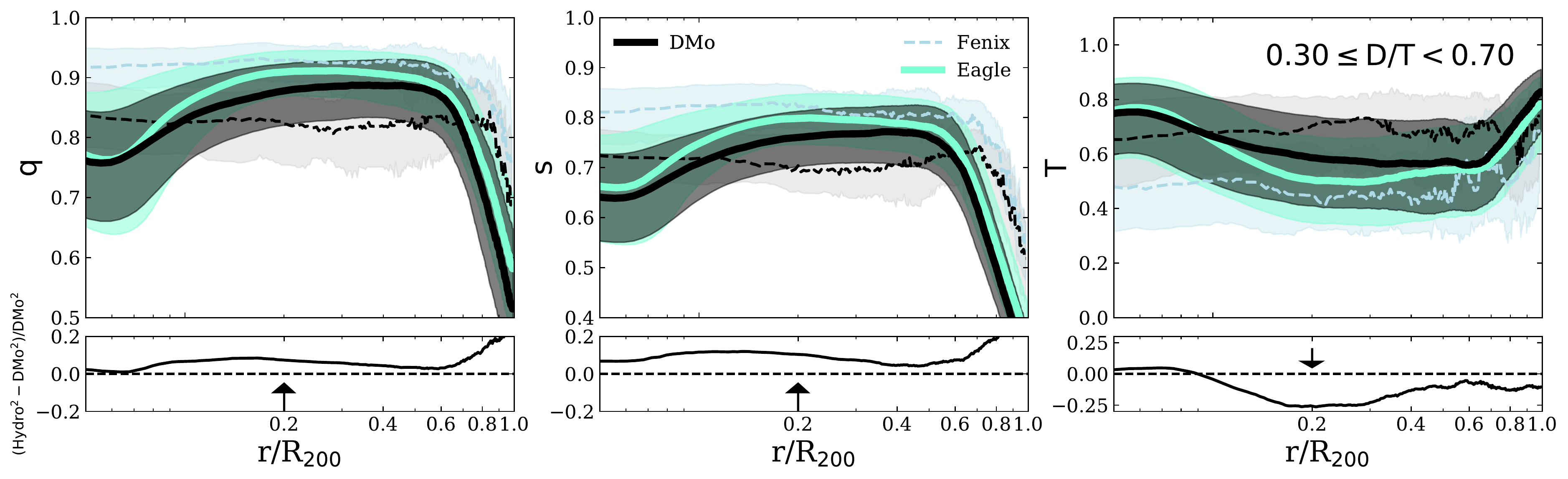}
  \includegraphics[width=\textwidth]{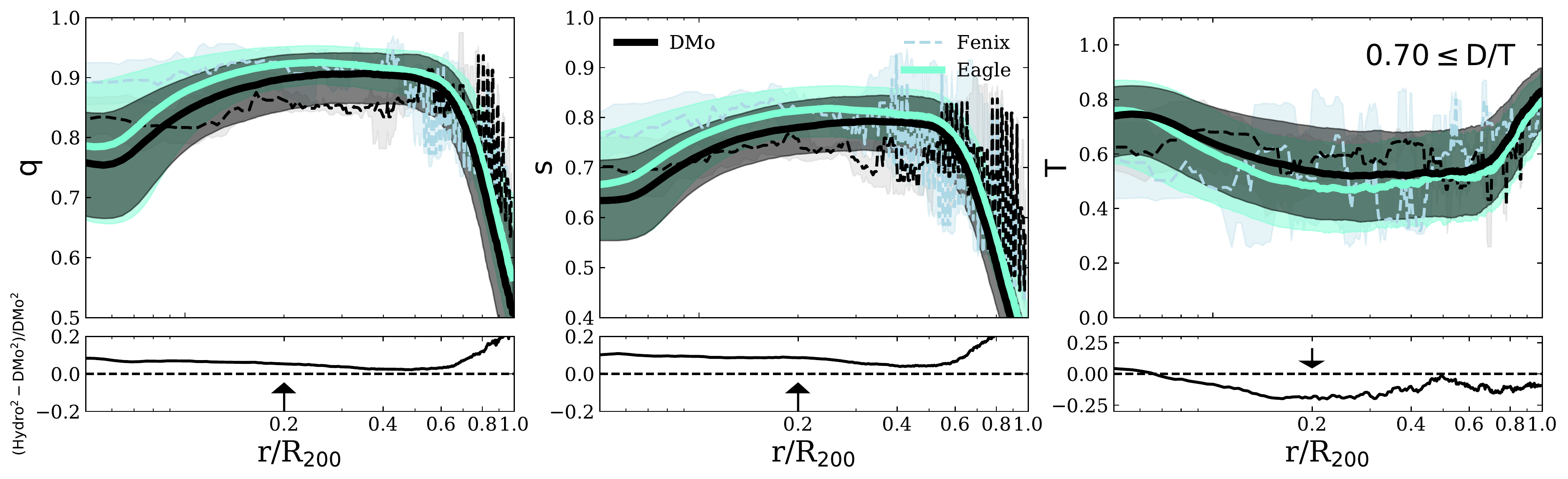}
\caption{Shape parameters $\rm q$, $\rm s$ and $\rm T$ median values versus $\rm r/R_{200}$ for {\sc eagle} (solid lines) and Fenix haloes (dashed lines), and for Hydro (light green) and its DMo counterpart (black) simulations. The shaded areas enclose the $25th$ and $75th$  quartiles. Haloes have been  divided in three subsamples according to the morphology of the hosted galaxies: $\rm D/T < 0.3$, $0.3 \leq \rm D/T < 0.7$ and $0.70 \leq \rm D/T$. Each panel includes the relative change between the shape parameters of the Hydro runs with their DMo counterparts for the {\sc eagle} subsample. The arrows indicate the radii that correspond to  20\% of the virial radius.}
\label{fig:fig3}
\end{figure*}

Fig.~\ref{fig:fig3} shows the shape parameters $\rm q$, $\rm s$ and $\rm T$ median values as a function of radius normalized by the virial one, $\rm r/R_{200}$,  for Hydro (light green) and DMo (black) runs. Three D/T intervals  are evaluated for both Fenix and {\sc eagle} simulations, $\rm D/T < 0.3$, $0.3 \leq \rm D/T<0.7$ and $0.7 \leq \rm D/T$ . The shaded areas enclose the $25th$ and $75th$ central quartiles. It can be seen that near the virial radius the effects of baryons on T diminish, recovering the triaxiality typical of the outer regions, in agreement with previous studies of \citet{Allgood} and  \citet{Butsky}. In the lower sub-panels, the relative change of the halo shape with respect to the DMo counterpart is quantified by defining  $\mathrm{(Hydro^{2} -DMo^{2})/DMo^{2}}$ (in the case of the s shape parameter, it means $\mathrm{(s_{Hydro}^{2} -s_{DMo}^{2})/s_{DMo}^{2}}$), for the {\sc eagle} haloes. The signal appears to maximize around  $\rm 0.20R_{200}$ independently of the predominance of stellar disc. For Fenix haloes subsample we do not detect a clear strengthening of the signal for a particular radius. The relative change of the median values of the shape parameters in Fenix haloes are greater than in {\sc eagle} haloes.  

Shape changes have been shown to be related with the modification of the orbital structure of the DM particles \citep{Zhu17}. Hence the region that maximizes the effects of the shape changes could be located outside the main galaxy, shifted with respect to the most inner region where haloes concentration changes are stronger.

In a previous study using the NIHAO zoom-in simulations, \citet{Butsky} choose  $\rm 0.12R_{ 200}$ as the characteristic radius, based on the findings of \citet{Ibata2001} of the Sagittarius stream location within 20-60~kpc. \citet{Thob}, using a similar subsample of {\sc eagle} haloes, use a fixed aperture of $\rm 30 \, kpc$ for the whole sample. As it comes out from  Fig.~\ref{fig:fig3}, the strength of the effects of the baryons on the DM haloes is sensitive to the regions where it is measured. Our results indicate that the effects on the shape are stronger in the range of 20--40 percent of the virial radius.

\begin{figure*}
  \centering
  \includegraphics[width=\columnwidth]{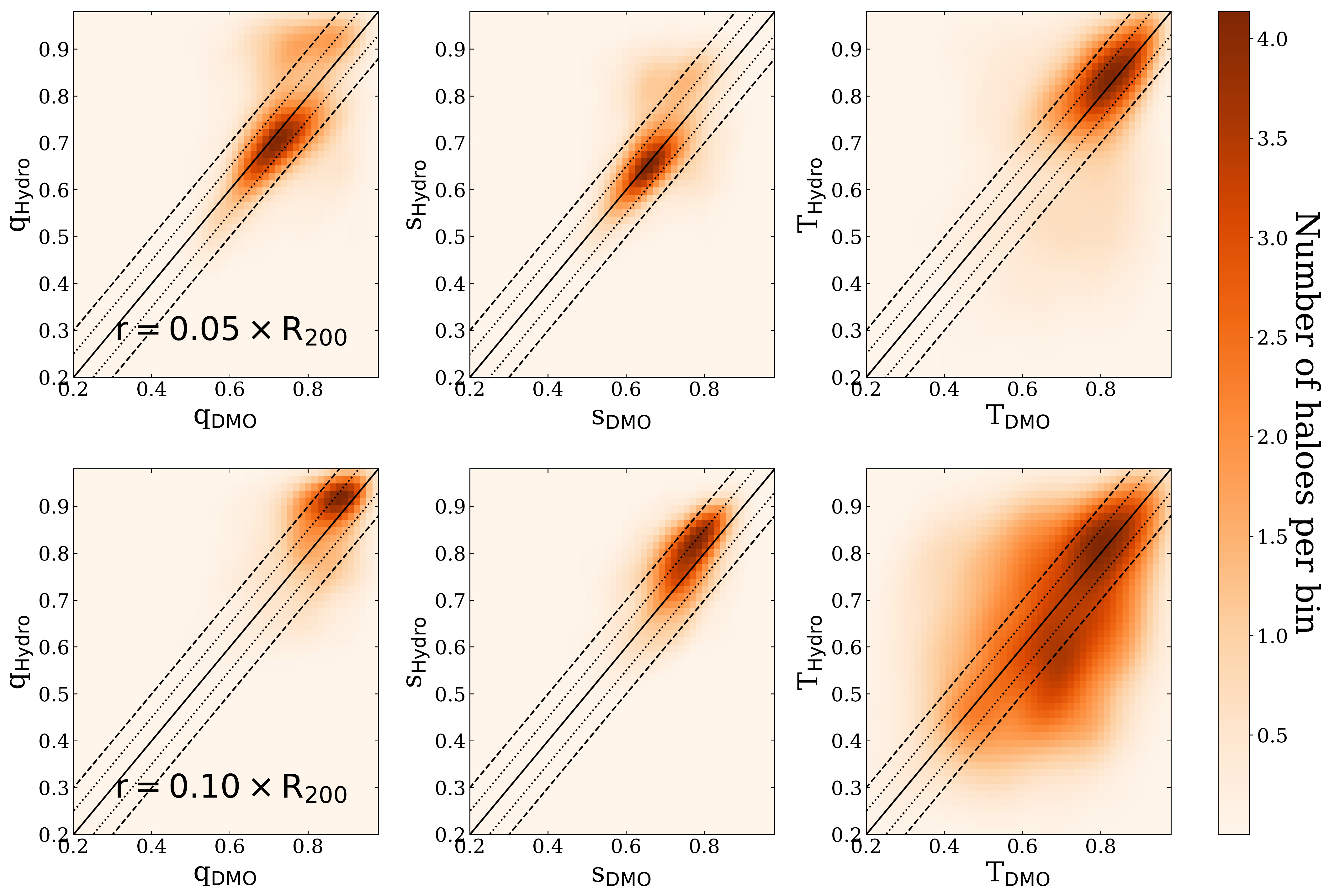}
  \includegraphics[width=\columnwidth]{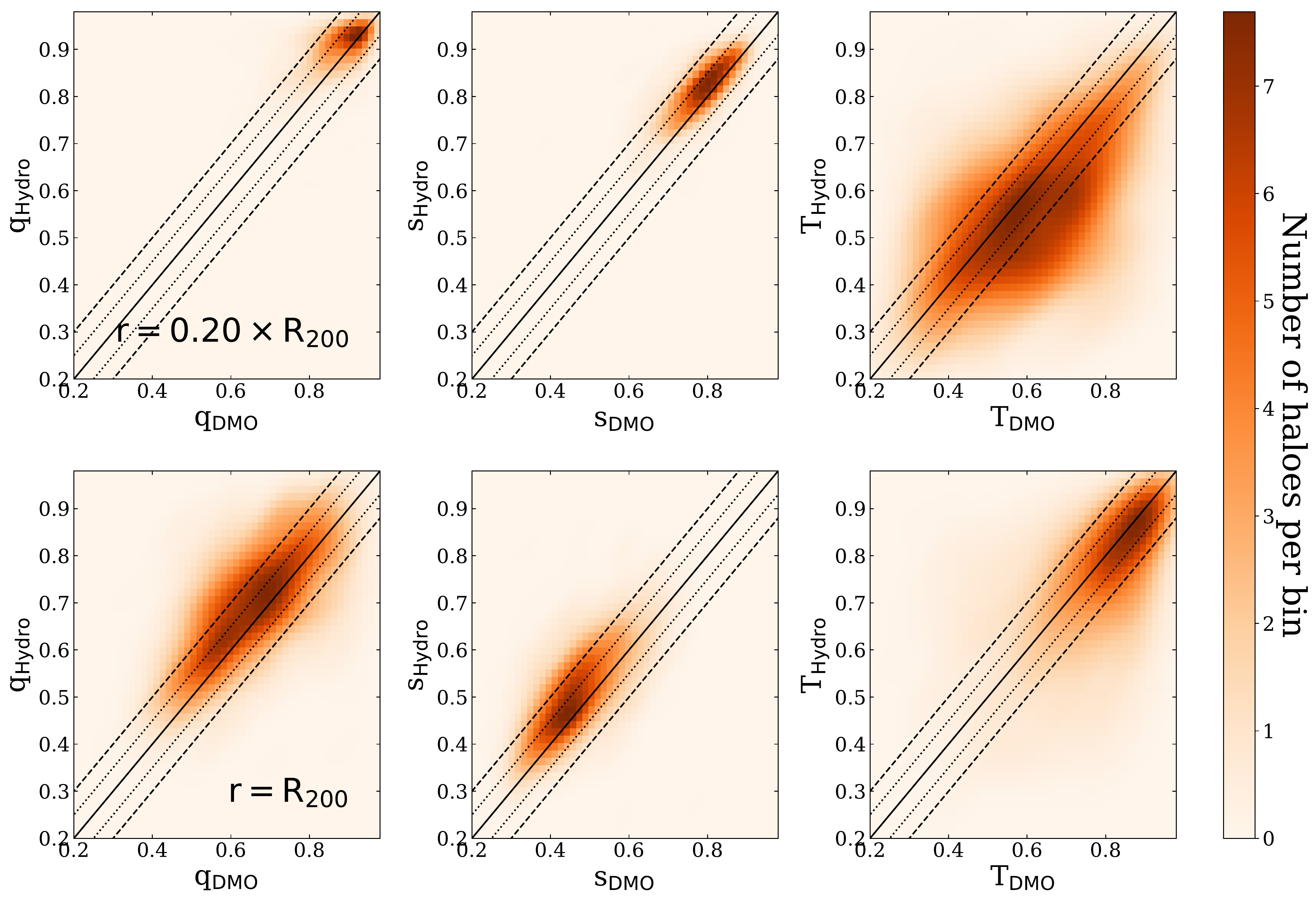}
\caption{$\rm 2D$ histogram  of the DM halo shape parameters for Hydro versus DMo simulations measured at $\rm r=0.05R_{\rm 200}$ (top left), $\rm r=0.10R_{\rm 200}$ (bottom left), $\rm r=0.20R_{\rm 200}$ (top right) and $\rm r=R_{\rm 200}$ (bottom right) for the {\sc eagle} haloes. The 1:1 relations are shown (solid lines) together with the 5 and 10 percent deviations (dotted lines) for both Hydro and DMo runs. The inner haloes from the Hydro run are more spherical and oblate than the DMo counterparts. At the virial radius, the effect of baryons is less significant whereas the maximum deviations are also found for $\rm r=0.20R_{\rm 200}$.}
  \label{fig:fig5}
\end{figure*}

Another way to visualize the modifications of the shape for the different radii can be appreciated in Fig.~\ref{fig:fig5} where we show $\rm 2D$ histograms comparing the shape parameters for Hydro and DMo haloes for the {\sc eagle} simulation at the four selected radii: $\rm r=0.05R_{\rm 200}$, $\rm r=0.10R_{\rm 200}$, $\rm r=0.20R_{\rm 200}$ and the virial one $\rm R_{\rm 200}$ \footnote{for the Fenix haloes see Fig.~\ref{fig:fig4} in the Appendix, Section \ref{sec:appendix}}. In the outer regions the $\rm s$, $\rm q$ and $\rm T$ parameters are almost identical for both runs as changes in halo shape are less significant near the virial radius. The dispersion for $\rm s$ and $\rm q$ is greater at this radii. In the inner regions, the largest deviation from the identity is detected at $\rm r=0.20R_{\rm 200}$ and is consistent with more triaxials haloes in the DMo run.These trends are in good agreement with the results found by \citet{Chua} analysing the Illustris  haloes. Our results indicate that the impact on the shapes are stronger close to  $\rm \sim 0.20 \, \rm R_{\rm 200}$.

\subsubsection{Shape dependence on mass}

\begin{figure}
  \centering
   \includegraphics[width=1.\columnwidth]{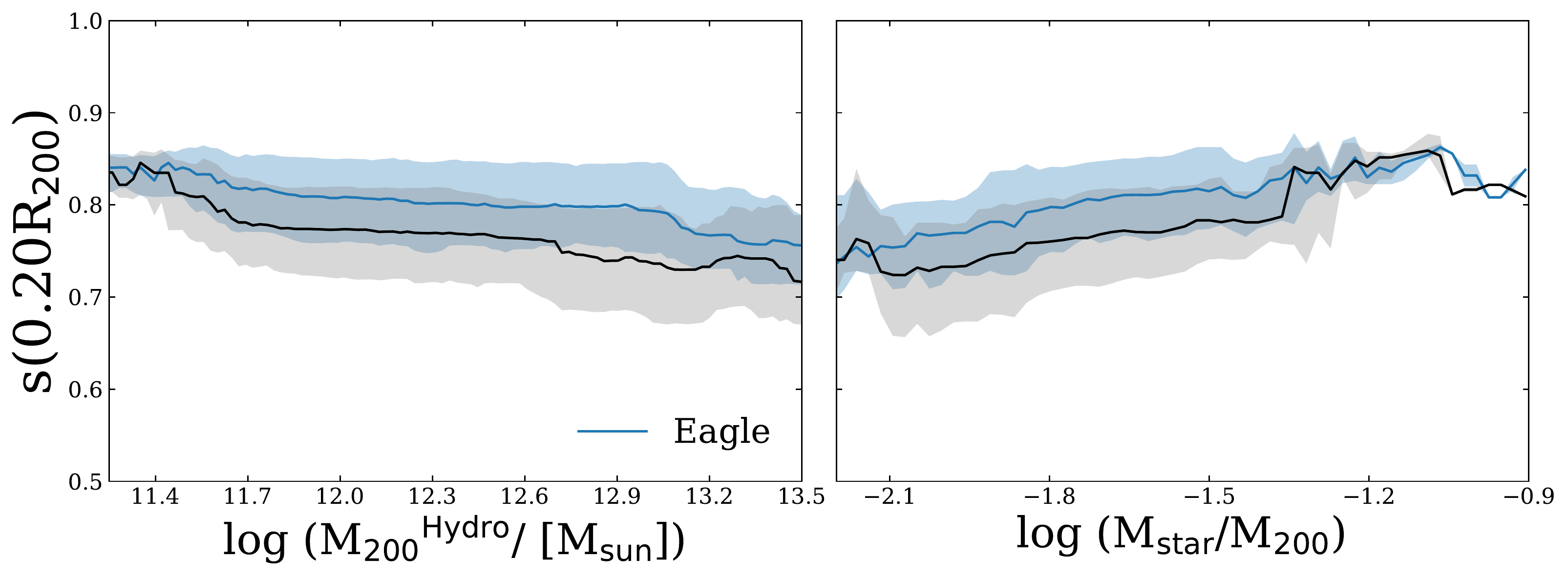}
   \includegraphics[width=1.\columnwidth]{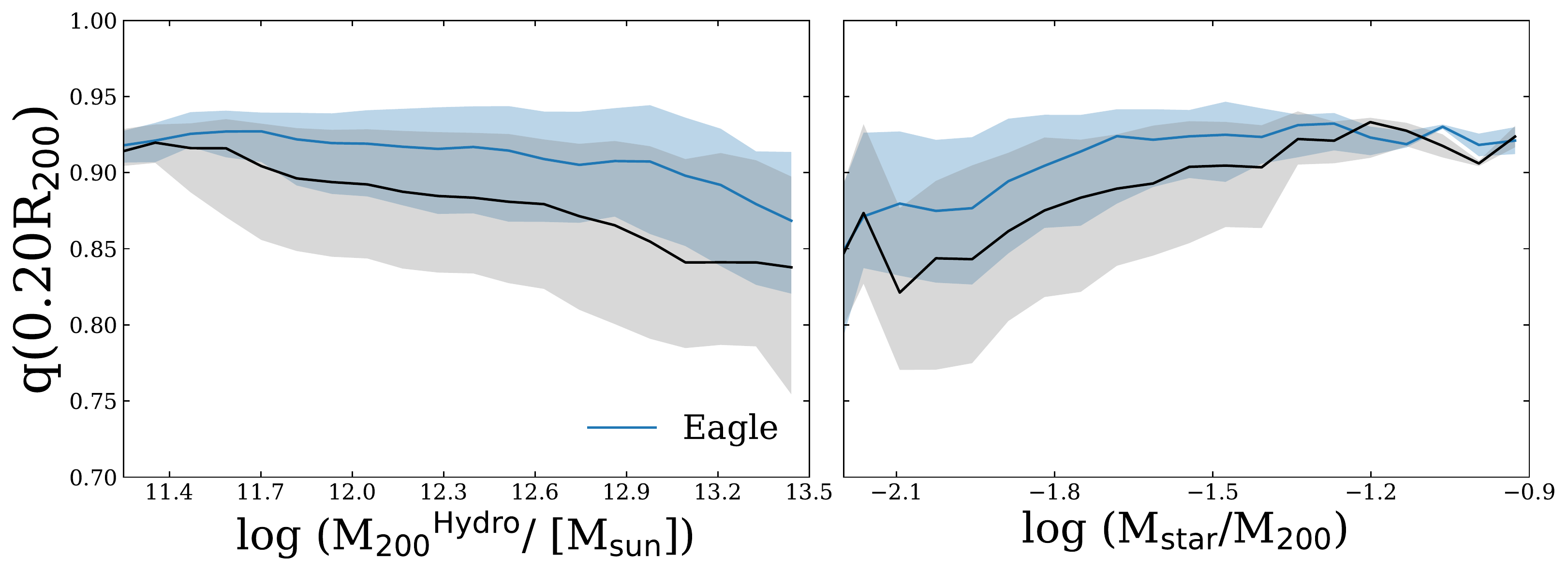}
   \includegraphics[width=1.\columnwidth]{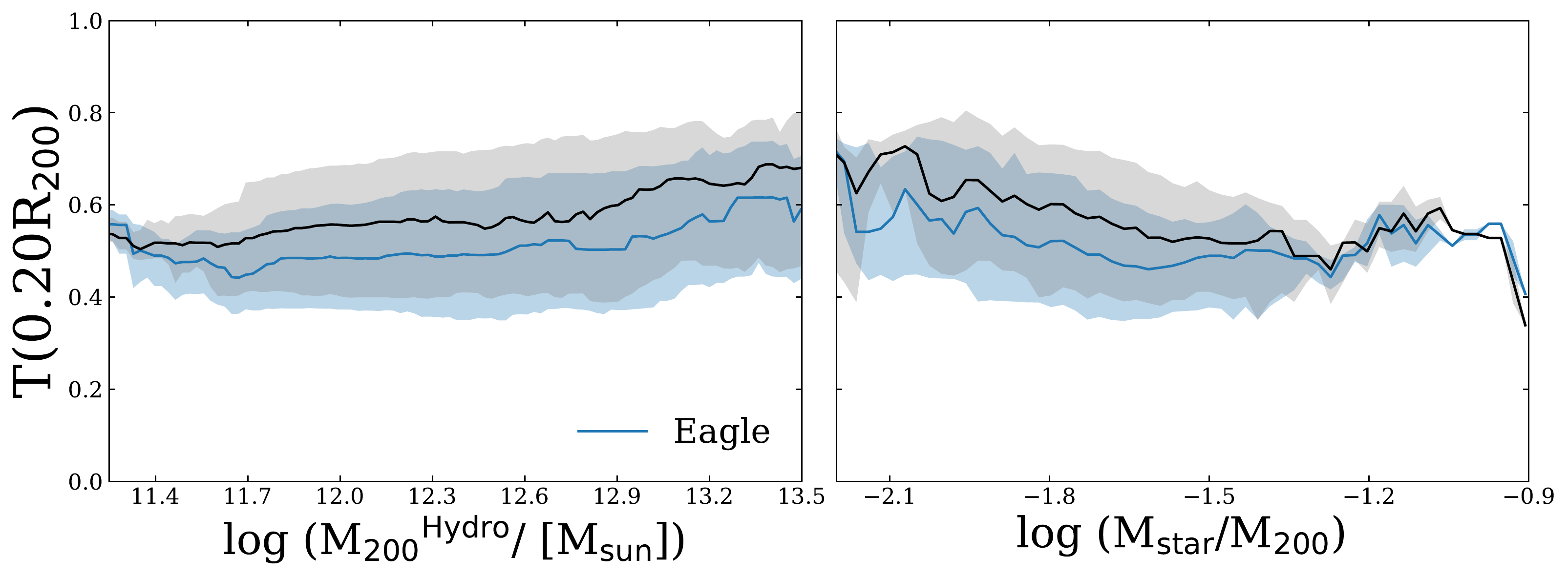}
  \caption{DM halo shape parameters $\rm s$, $\rm q$ and $\rm T$, for the {\sc eagle} simulation, measured  at 20 percent of $\rm R_{200}$ as a function of  $\rm M_{200}^{\rm Hydro}$ (left panels) and the stellar mass - halo mass ratio, $\rm M_{\rm star}/M_{\rm 200}$ (right panels). Solid lines indicate the median of the distributions for Hydro (light blue) and DMo (black) haloes, while the shaded areas enclose the $25th$ and $75th$ central quartiles. For higher masses $\rm M_{200}^{\rm Hydro}$ here is a trend for haloes to be less spherical. Haloes with higher $\rm M_{\rm star}/M_{\rm 200}$  are more spherical.}
  \label{fig:fig6}
\end{figure}

We also explore the mass dependence of the shape parameters. We inspect this with respect to both, the stellar mass fraction $\rm M_{star}/M_{200}$ and $\rm M_{200}$ at 20 percent of $R_{200}$. Fig. \ref{fig:fig6} shows the shape parameters as a function of the virial mass $\rm M_{200}$ and the ratio $\rm M_{star}/M_{200}$. For both Hydro and DMo runs, more massive haloes tend to be more triaxials. On the other hand, when considering the dependence on $\rm M_{star}/M_{200}$, the triaxiality gets lower for larger stellar mass as expected \citep{Butsky,Chua}.

\begin{figure}
\centering
  \includegraphics[width=\columnwidth]{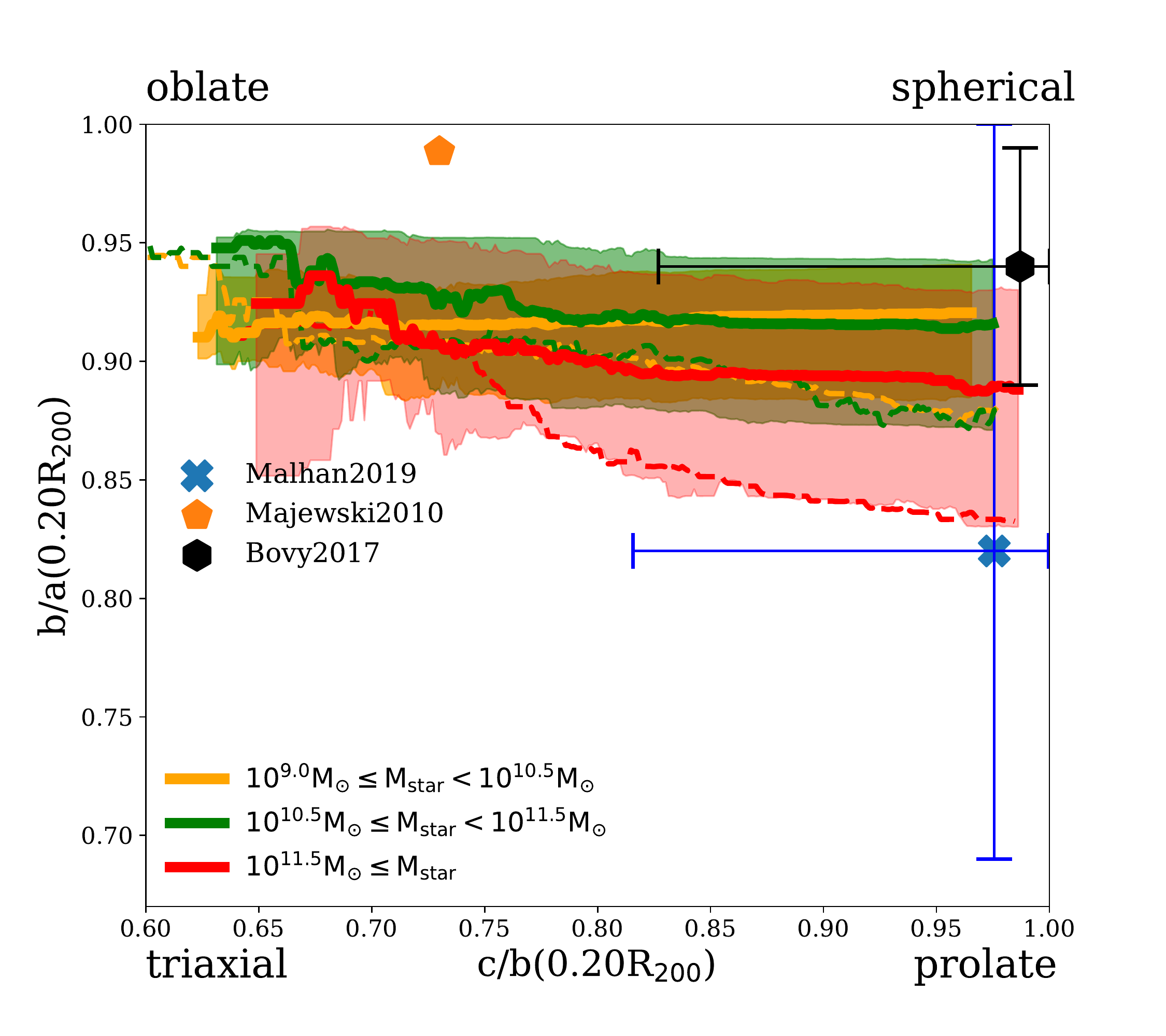}
    \caption{The distribution of {\sc eagle} haloes as a function of their inner halo axial ratios measured at 20 percent of the virial radius for different $\rm M_{star}(0.20R_{200})$. In the upper right corner, when $\rm b/a \sim   1.0$ and $\rm c/b \sim   1.0$, the haloes are more spherical. The shaded areas denote the $\rm 25th$ and $\rm 75th$  quartiles. Dashed lines indicate the median of the distribution for the correspondent DMo counterpart. In blue, orange and black symbols the observational constrains for the DM halo shape in the Milky Way by \citet{Law2010}, and \citet{Bovy2016}, \citet{Malhan}.}
\label{fig:fig7}
\end{figure}

To better visualize these trends, Fig.~\ref{fig:fig7} shows, for the {\sc eagle} haloes, the axial ratios $\rm b/a$ as a function of $\rm c/b$ within $\rm 0.20R_{\rm 200}$ for different intervals of $\rm M_{\rm star}$. In this plane, haloes close to the top right corner are more spherical ($\rm a\sim 1$, $\rm b\sim 1$, $\rm c\sim 1$) as described in \citet{Trayford}. 

There is a rather statistically weak trend for more prolate halo shapes to be found hosting more massive stellar objects. \citet{Butsky} found a strong mass dependence for the NIHAO zoom-ins: low mass haloes tend to retain their original triaxiality while for higher mass haloes, the inner regions  become more spherical.  

Fig.~\ref{fig:fig7} also shows observational constrains for the DM halo shape of the Milky Way at different radius, spanning between  $\sim 20-40$~kpc. Our results show good agreement with \citet{Bovy2016} and \citet{Malhan}. The value found by \citet{Law2010} is higher than ours, and are significantly different between each other. The best fitting \citet{Law2010} model has an oblate halo, with axes ratios, $\left \langle c/a \right \rangle=0.72$ and $\left \langle b/a \right \rangle=0.99$.

Fenix and {\sc eagle}  cover different stellar mass ranges mainly due to the limited size of the Fenix simulation box. Despite the different feedback implementation both trends present a general agreement, within the same mass range. On the other hand, haloes from {\sc eagle} simulation have a wider stellar mass range and morphology because of the small volume which includes a rich variety of environments: field, groups and clusters (Fenix represents a field region only). 

\section{Shape Dependence on  morphology}\label{sec:shapemorphology}

Previous numerical studies find that haloes become rounder when they host baryons in their centres, and more recent ones suggest that it might exist a correlation with the shape of the galaxy (see \citealt{Tissera1998}, for an earlier comment on this aspect).

\begin{figure*}
\centering
  \includegraphics[width=\columnwidth]{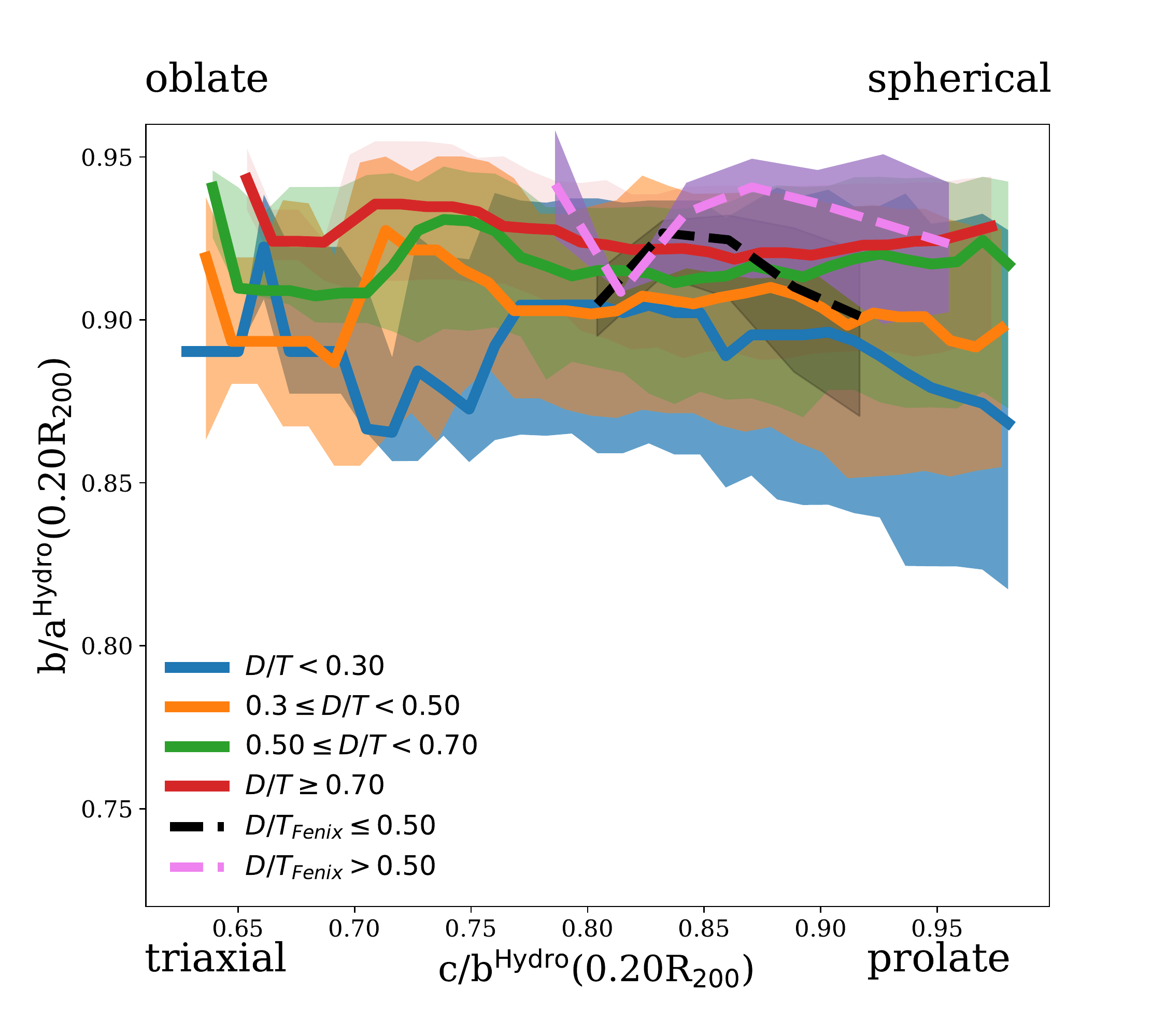}
  \includegraphics[width=\columnwidth]{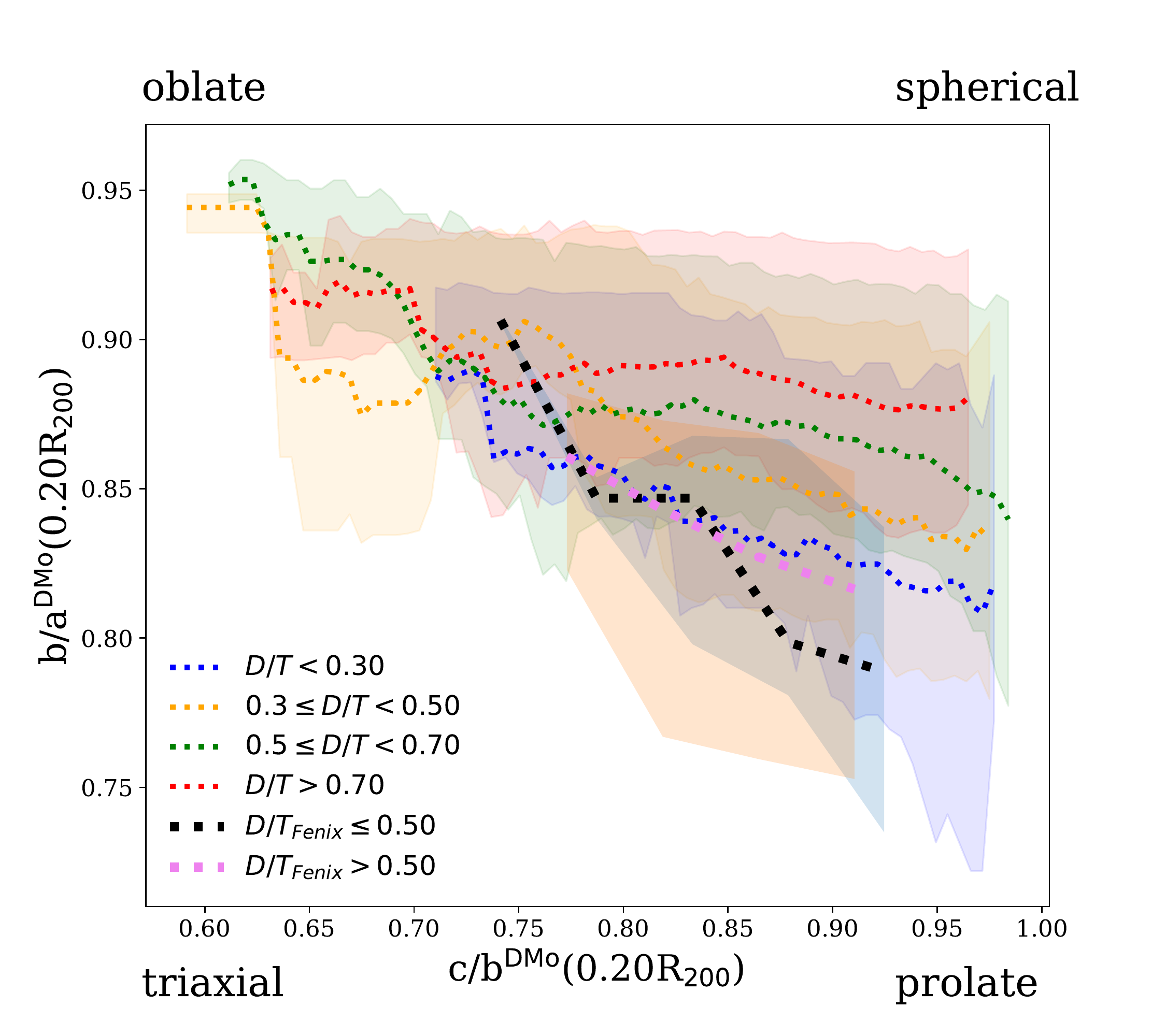}
    \caption{The distribution of {\sc eagle} and Fenix for the Hydro selected haloes (left panel) and their DMo counterparts (right panel) as a function of their inner halo semi-axis ratios measured at 20 percent of the virial radius. The relations are given for galaxies with different morphologies quantified by the $\rm D/T$ ratios. On the left panel, solid (dashed) lines indicate the median of the distribution for the adopted $\rm D/T$ intervals for {\sc eagle} (Fenix) galaxies. Right panel shows the same for the DMo counterparts. The shaded areas denote the $\rm 25th$ and $\rm 75th$  quartiles.}
\label{fig:fig8}
\end{figure*}

Recently, \citet{Thob} found that the morphology of {\sc eagle} galaxies present a slight correlation with the morphology of its host halo. Furthermore, they  suggest that this correlation could have an intrinsic rather than induced origin. In order to further inspect this, in Fig.~\ref{fig:fig8} we show the semi-axis ratios as a function of the $\rm D/T$ ratio, for the Hydro haloes (left panel) and their DMo counterparts (right panel) at $\mathrm{0.20R_{200}}$ (for the same plot estimated at 5 and 10 percent of $\rm R_{200}$ see Fig. \ref{fig:fig12} and Fig.\ref{fig:fig13} in Appendix, Section \ref{sec:appendix}).

There is a trend for haloes at $\mathrm{0.20R_{200}}$ to be globally more oblate for higher $\rm D/T$ ratios in both {\sc eagle} and Fenix simulations (although in the last one, only two $\rm D/T$ intervals could be defined due to the smaller sample). Our results are in good agreement with \citet{Kazantzidis}.  A clear intrinsic trend in the DMo shapes with the $\rm D/T$ fraction can be appreciated.

The presence of baryons alters the DM haloes by making them less triaxials. Some authors suggest that this effect is enhanced when the baryons are organized in disc structures \citep{Kazantzidis}. However,  the  trend that can be seen in Fig.~\ref{fig:fig8} for the DMo runs also suggest that  extended disc galaxies preferentially form in haloes that are intrinsically more spherical (i.e. the DMo counterparts are more spherical). 

The connection between the inner DM halo and the characteristics of the galaxy it can host was previously mentioned by \citet{Zavala} by studying the angular momentum evolution of the galaxies in relation with their DM haloes and by \citet{Thob} who analyse the galaxy flattening in comparison the flattening on the DM haloes. Both works used the {\sc eagle} simulation for this purpose. We made a step forward and measured the degree of sphericalization for different morphologies and quantify with the  $\rm D/T$ fraction in relation with the DMo shapes. 

Our results show that the combination of the DMo halo masses, used as an input in semianalytical models, and shapes could be use to predict the probability that a halo hosts a galaxy with a specific morphology.

\begin{figure}
  \centering
  \includegraphics[width=0.225\textwidth]{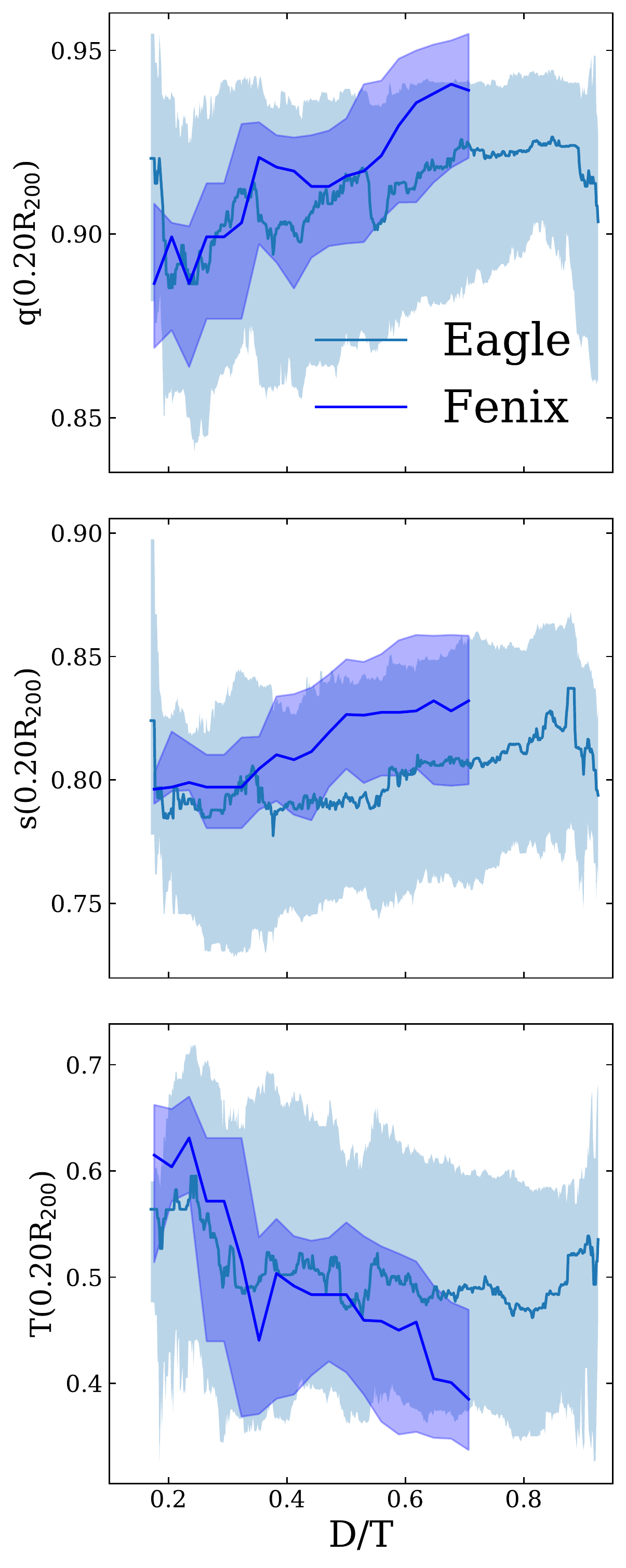}
  \includegraphics[width=0.22\textwidth]{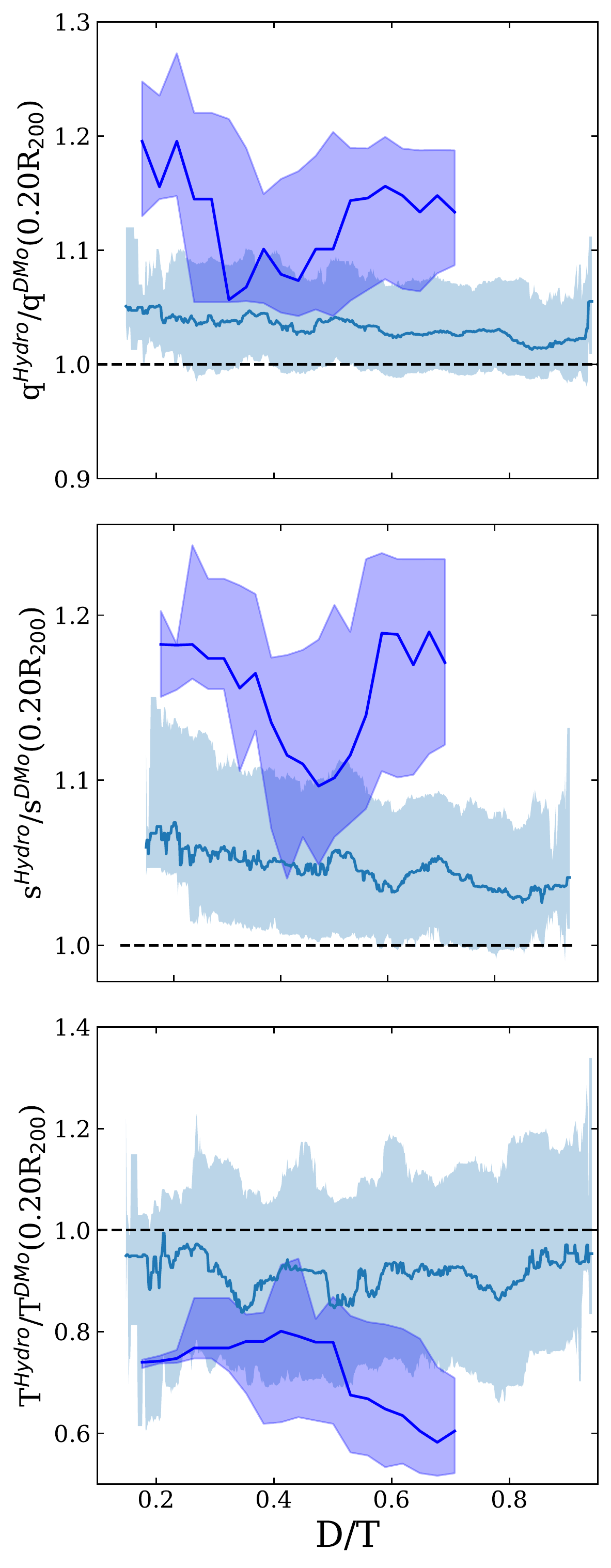}
  \caption{DM halo shape parameters $\rm q$,  $\rm s$, $\rm T$  measured at the 20 percent of the virial radius in the Hydro runs as a function of $\rm  D/T$ (left column). Right panel shows the shape parameters of the Hydro, normalized by those from their DMo  counterpart. Solid  lines indicate the median of the distribution for {\sc eagle} (light blue) and Fenix haloes (blue), respectively.  The shaded areas enclose the $25th$ and $75th$  quartiles.  We find that haloes tends to be more rounder and less triaxials for larger $\rm D/T$. The correlation become weaker when normalized by the DMo counterparts.}
  \label{fig:fig9}
\end{figure}

In order to better display this trend, in Fig \ref{fig:fig9} we show the DM halo shape parameters $\rm q$,  $\rm s$, $\rm T$,  measured at the 20 percent of the virial radius, as a function of the $\rm D/T$. For the Fenix haloes the D/T ratios cover a  shorter range  ($\mathrm{D/T_{\, Fenix} \: \epsilon \: \left [ 0.16, 0.73 \right ]}$), as previously mentioned, while for the  {\sc eagle} haloes larger D/T ratios are found ($\mathrm{D/T_{\, EAGLE}  \: \epsilon \: \left [ 0.15, 0.95 \right ]}$ ). 

\begin{table}
\centering
\caption{\label{tab1} The Spearman, $\rho$, and Pearson, r, correlation coefficient (with their correspondent p-values), for the shape parameters measured within the 0.20R$_{200}$ as a function of $\rm D/T$ ratios. The correlation weakens when the parameters are normalized by their correspondent DMo counterparts.}
\resizebox{\columnwidth}{!}{%
\begin{tabular}{llll}
\hline
\\[-5pt]
 Spearman $\, \rho $ (p-value) & $\rm q^{\rm Hydro}$ & $\rm s^{\rm Hydro}$ & $ \rm T^{\rm Hydro}$ \\
\hline
$\mathrm{D/T_{\, Fenix} } $    & $0.43 \, (7x10^{-3})$ & $0.23  \, (0.16)$ & $-0.33 \, (0.04)$ \\

$\mathrm{D/T_{\,  {\sc EAGLE}} } $  & $0.14 \, (1x10^{-8})$  & $0.13 \, (4x10^{-8})$ & $-0.08 \, (8x10^{-4})$ \\

\hline
\\[-5pt]
 Pearson r (p-value) & $\rm q^{\rm Hydro}$ & $\rm s^{\rm Hydro}$ & $ \rm T^{\rm Hydro}$ \\
\hline
$\mathrm{D/T_{\, Fenix} } $    & $0.39 \, (0.01) $ & $0.21 \, (0.22)$ & $-0.34 \, (0.04)$ \\

$\mathrm{D/T_{\,  {\sc EAGLE}} } $  & $0.15 \, (5x10^{-10})$ & $0.13 \, (9x10^{-8})$ & $-0.09 \, (2x10^{-4})$ \\

\hline
\\[-5pt]
  Spearman $\,\rho$  (p-value) & $ \rm q^{\rm Hydro}/ \rm q^{\rm DMo}$ & $\rm s^{\rm Hydro}/\rm s^{\rm DMo}$ & $\rm T^{\rm Hydro}/\rm T^{\rm DMo}$ \\
\hline
$\mathrm{D/T_{\, Fenix} } $    & $-0.04 \, (0.79)$ & $0.04 \, (0.81)$ & $-0.24 \, (0.15)$ \\

$\mathrm{D/T_{\,  {\sc EAGLE}} } $  & $-0.07 \, (3x10^{-3})$ & $-0.09 \, (1x10^{-5})$ & $0.08 \, (0.65)$ \\

\hline
\\[-5pt]
  Pearson r (p-value) & $ \rm q^{\rm Hydro}/ \rm q^{\rm DMo}$ & $\rm s^{\rm Hydro}/\rm s^{\rm DMo}$ & $\rm T^{\rm Hydro}/\rm T^{\rm DMo}$ \\
\hline
$\mathrm{D/T_{\, Fenix} } $    & $-0.14\, (0.38)$ & $-0.03 \,(0.86)$ & $-0.14\, (0.42)$ \\

$\mathrm{D/T_{\,  {\sc EAGLE}} } $  & $-0.09\, (4x10^{-4})$ & $-0.10 \, (6x10^{-5})$ & $0.02 \, (0.34)$ \\

\hline
\end{tabular}%
}
\end{table}

In the left panels of Fig \ref{fig:fig9}, it can be appreciated clearly that haloes are rounder and less triaxials for larger  $\rm D/T$. However,  when this relation is normalized by their correspondent DMo counterpart, the correlations strongly weakens as shown on the right panel. We estimate the Spearman and Pearson correlation coefficients in Table~\ref{tab1}. The p-values of Fenix sample show a greater deviation from 0 what impact negatively in the reliability for the strength of the correlation. This is to be expected since the small size of the sample impact on the p-values.

This supports a cosmological origin for the link between DM haloes and the morphology of their central galaxies, which can be also affected by other processes during its assembly history, producing a large dispersion in the relation.

Recall that there is a dependence of the shape parameters on $\rm M_{star}$ and $\rm M_{star}/M_{200}$. The relations shown in this figure consider all haloes regardless of their stellar content. We note that Fenix haloes cover the lower range of stellar masses of {\sc eagle} and hence, they tend to be rounder compared to the overall median values.

\subsection{DM velocity structure}

The velocity anisotropy of individual haloes presents a variety of behaviours depending on their particular formation history \citep[see e.g.,][]{Tissera2010}. In this section we analyse the change in velocity anisotropy of the DM haloes  as a function of radius and the morphology of the hosted galaxy. The velocity structure of the DM particles are closely related with the resulting shapes of the haloes. Additionally, it has important implications for the predicted scattering rates in direct detection experiments \citep{Kuhlen2010}.

We inspect the velocity anisotropy parameter defined as:

\begin{equation}
\beta (r) = 1- \frac{\sigma _{t}}{2\sigma _{r}},
\end{equation}
\noindent  
 
where $\sigma _{r}$ and  $\sigma _{t}$ are the radial and tangential velocity dispersions averaged over concentric spherical shells. This anisotropy parameter provides some indication of the velocity distribution of the haloes. Isotropic distributions have $\beta \approx 0$ whereas those that more radially biased have $\beta > 0$.

\begin{figure}
\centering
  \includegraphics[width=\columnwidth]{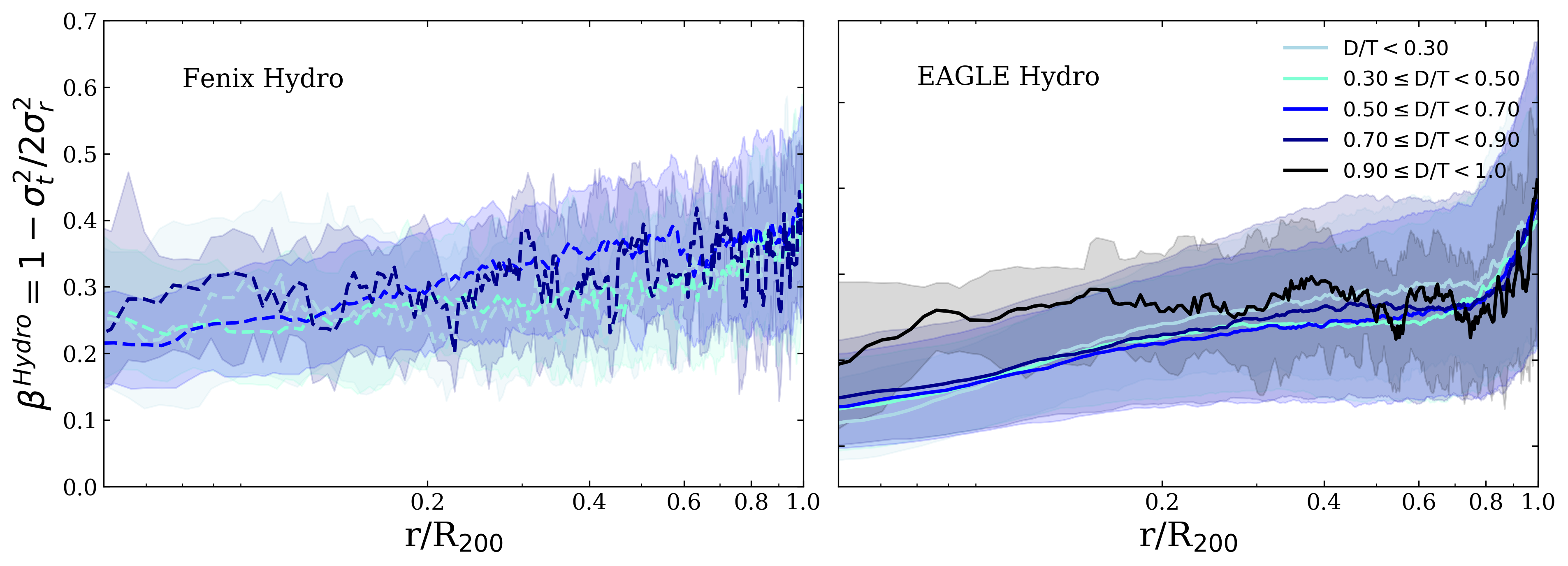}
  \includegraphics[width=\columnwidth]{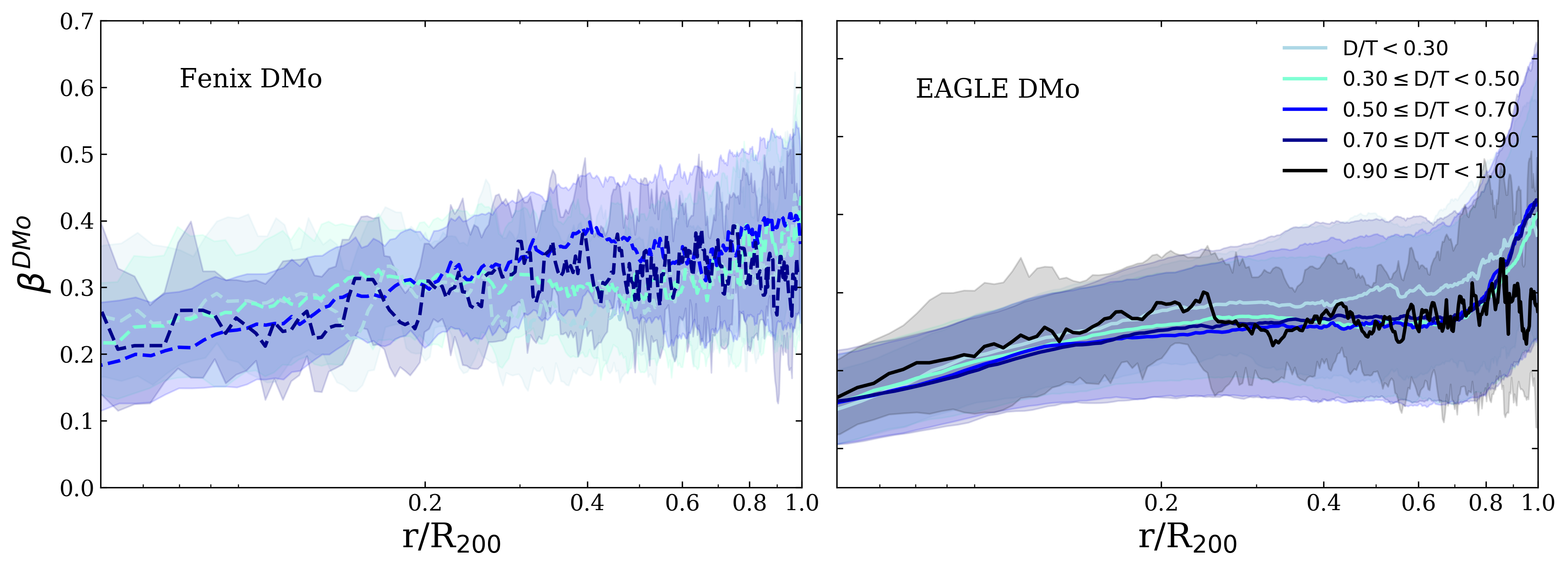}
\caption{ $\rm \beta$  as a function of $\rm r/R_{\rm 200}$.  Dotted lines indicate the median of the distribution in the case of the Fenix haloes (left panel) and solid lines in the case of the {\sc eagle} haloes (right panel). The $\beta$  is divided with  disc fraction bins  $\rm D/T<0.30$, $0.30 \leq \rm D/T < 0.50$ and $0.50 \leq \rm D/T  < 0.70$. $0.70 \leq \rm D/T  < 0.90$ and $ 0.90  \leq \rm D/T $.}
\label{fig:fig10}
\end{figure}

In Fig. \ref{fig:fig10} we show the median distributions of $\beta (r)$ as a function of the radius normalized by the virial one, $\rm r/R_{\rm 200}$, for galaxies grouped according to their morphology: $\rm D/T<0.30$, $0.30 \leq \rm D/T < 0.50$, $0.50 \leq \rm D/T  <  0.70$, $0.70 \leq \rm D/T  < 0.90$, and $ 0.90  \leq \rm D/T $, for the Fenix (left panel) and {\sc eagle} (right panel) haloes. Both {\sc eagle} and Fenix simulations show a noticeable trend  for radius smaller  than approximately $\mathrm{0.20R_{200}}$: haloes hosting almost bulgeless galaxies present higher values of $\beta$. The velocity structure within the central regions of DM haloes hosting the highest D/T fraction depart from  isotropy. For radius larger than $\rm \sim 0.20R_{200}$ this trend disappear.

When the DMo is inspected (lower panels), $\beta ^{DMo} (r)$ show a slight trend in the same sense albeit weaker. As we have found for the shapes, there is an intrinsic behaviour indicating that haloes that host important stellar disc structures tend to have a distinct, slightly less isotropic, DM velocity pattern. To better quantify this behaviour, we estimate $(\beta - \bar{\beta})/\bar{\beta}$  as a function of $\rm r/R_{\rm 200}$ for both the Hydro and DMo runs for galaxies with different morphologies.  

\begin{figure}
\centering
  \includegraphics[width=0.7\columnwidth]{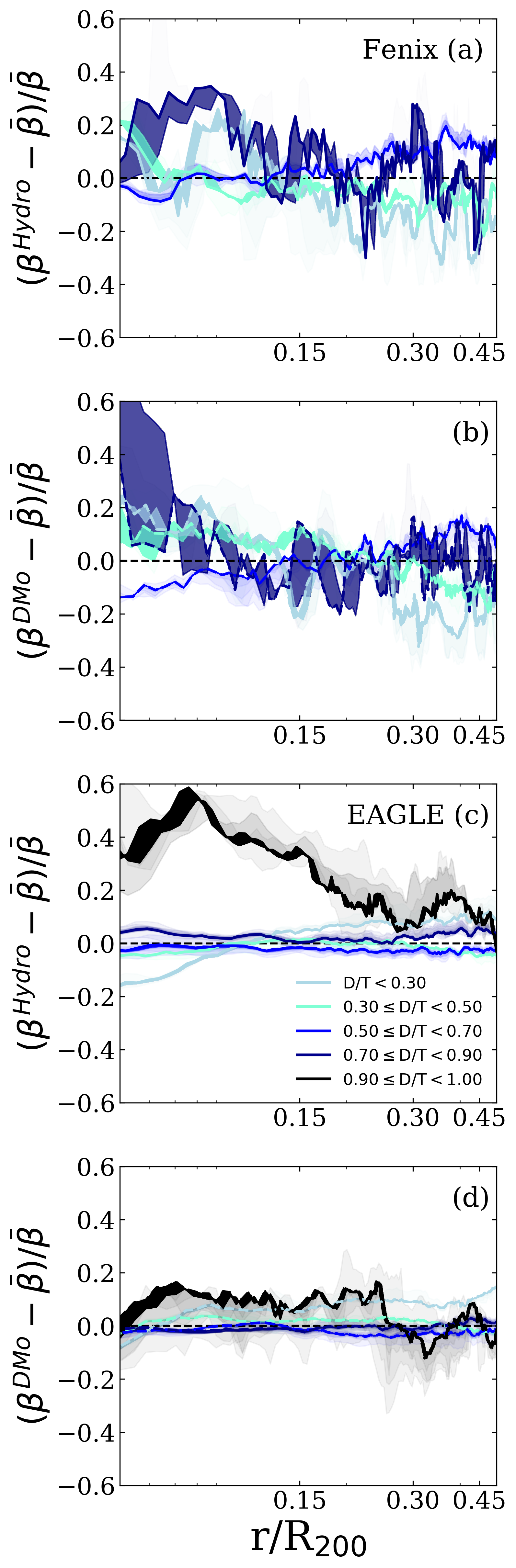}
\caption{ $(\beta - \bar{\beta})/\bar{\beta}$  as a function of $\rm r/R_{\rm 200}$ for Fenix Hydro and DMo haloes (panels a,b) and {\sc eagle} Hydro and DMo haloes (panels c,d). Errors were computed using a bootstrap method and each shaded areas denote the number of computed bootstrap sample. Haloes with higher $\rm D/T$ present a deviation from the isotropic distribution in the inner regions. DMo haloes have the same trend.}
\label{fig:fig11}
\end{figure}

Fig. ~\ref{fig:fig11} show the relative change of $\beta (r)$ with respect to the median relations (i.e $(\beta - \bar{\beta})/\bar{\beta}$) for both Fenix (panels a,b) and {\sc eagle} (panels c, d) haloes as a function of $\rm r/R_{\rm 200}$. $\beta ^{DMo} (r)$, panels b and d, show a slight trend in the same sense albeit weaker. As we have found for the shapes, there is an intrinsic behaviour indicating that haloes that host more important stellar disc structures tend to have a distinct, slightly less isotropic, DM velocity pattern. The error regions were computed by using a bootstrap method with a resampling of 200 for each D/T interval. As can be seen, the larger velocity anisotropies are found in haloes with almost bulgeness galaxies within $\sim \rm 0.20R_{200}$. The correspondent DMo counterpart haloes show similar estimations but using the $\beta^{\rm DMo}$.

A greater velocity anisotropy  could be related with modification in the orbit structure of DM velocity. We have also looked for DM discs in the central regions of the selected Fenix haloes without any detection.   \citet{Schaller2016} concludes that the presence of a dark disc is unlikely in the {\sc eagle} haloes. The DM configuration could be related to more tube-type orbits, as found by \citet{Zhu17}. So this change in orbit configuration towards higher central velocity anisotropies together with intrinsic DMo haloes shapes could foster preferentially the formation of stellar discs.

The presence of a similar excess in the same radial range is consistent with the claim of an intrinsic characteristics of the haloes that might contribute to the formation of the disc-dominated galaxies.

\section{Conclusions}
\label{sec:conclusions}

We investigate the relation between the properties of the DM haloes and the morphologies of the hosted galaxies. For this purpose we analysed  two simulations with different subgrid physics implementation: a subsample from from the  largest volume run of the {\sc eagle} Project (L100N1504) and haloes from  the Fenix Project. We investigate the effects on the halo shape in the central regions. Our main findings can be summarized as follow:
 
\begin{enumerate}

    \item{We find that haloes hosting galaxies with higher D/T fractions are rounder and less triaxials. Remarkably this trend is already present in the DMo haloes as shown in the right panel of Fig. \ref{fig:fig8}. This suggests that discs will form preferentially in haloes that are already more spherical showing that galaxy morphology is interlinked with the intrinsic structure of DM haloes (without baryonic physics). This could allow us to roughly statistically  predict the morphology a galaxy will have if it were to form in a halo of a given shape. This could be important for semi-analytical models.}
    
     \item{More massive haloes tend to be more triaxials. When considering the dependence on $\rm M_{star}/M_{200}$, the triaxiality gets lower for larger stellar mass as expected.}
     
     \item{We find that the impact of baryons  on the DM halo structure  maximizes when measured within \( \sim \rm 0.20 R_{200} \). While more compact galaxies are located in more concentrated DM haloes, the effects of galaxy assembly are clearer when measured within \( \sim \rm 0.20 R_{200} \). As expected, baryonic physics affects halo shape mainly in the inner regions. Haloes hosting baryons become less triaxials, in agreement with previous studies (see Fig. \ref{fig:fig3} and Fig. \ref{fig:fig5}).}
    
    \item{The analysis of the velocity structure of haloes using the anisotropy parameter shows a trend in the sense that almost bulgeless galaxies are associated with greater deviations from isotropy in the velocity distribution. This trend seems also to be intrinsic in the sense that it is also detected when using the velocity anisotropies of the DMo counterparts albeit weaker.}

    Our results shows that DM halo shapes, measured within $\sim \rm 0.20R_{200}$, are correlated with galaxy morphology. This correlations appears to be intrinsic as we detect that the same trend is already present in the DMo haloes. Hence, there is an indication that more important disc preferentially form in more rounder haloes. This intrinsic trend is reinforced by the growth of the disc galaxies, which strengths its cosmological origin. These results could potentially provide a statistical predictor of the morphology the galaxy would have if it were to form in a halo of a given shape.

\end{enumerate}

\section*{Acknowledgements}

We would like to thank the anonymous referee for the thorough revision of our  manuscript. The authors thank Matthieu Schaller for assistance with the DMo EAGLE data. PBT acknowledges partial support from Fondecyt Regular 1200703 and CONICYT project Basal AFB-170002(Chile). This project has received funding from the European Union’s Horizon 2020 Research and Innovation Programme under the Marie Skłodowska-Curie grant agreement No 734374 and the GALNET Network (ANID, Chile). Also this project was supported through PIP
CONICET 11220170100638CO; SP acknowledges partial support by  the Ministerio de Ciencia, Innovación y Universidades (MICIU/FEDER) under research grant PGC2018-094975-C21. MCA acknowledges financial support from the Austrian National Science Foundation through FWF stand-alone grant P31154-N27. 
This work used the DiRAC Data Centric system at Durham University, operated by the Institute for Computational Cosmology on behalf of the STFC DiRAC HPC Facility (www.dirac.ac.uk). This equipment was funded by BIS National E-infrastructure capital grant ST/K00042X/1, STFC capital grants ST/H008519/1 and ST/K00087X/1, STFC DiRAC Operations grant ST/K003267/1 and Durham University. DiRAC is part of the National E-Infrastructure.

\section*{Data Availability}

The data underlying this article will be shared on reasonable request to the corresponding authors.



\bibliographystyle{mnras}
\bibliography{Cataldietal} 




\appendix

\section{Extended analysis of halo shape}
\label{sec:appendix}

\setcounter{table}{0}
\renewcommand{\thetable}{A\arabic{table}}

\setcounter{figure}{0}
\renewcommand{\thefigure}{A\arabic{figure}}

\begin{figure*}
\centering
\includegraphics[width=0.97\columnwidth]{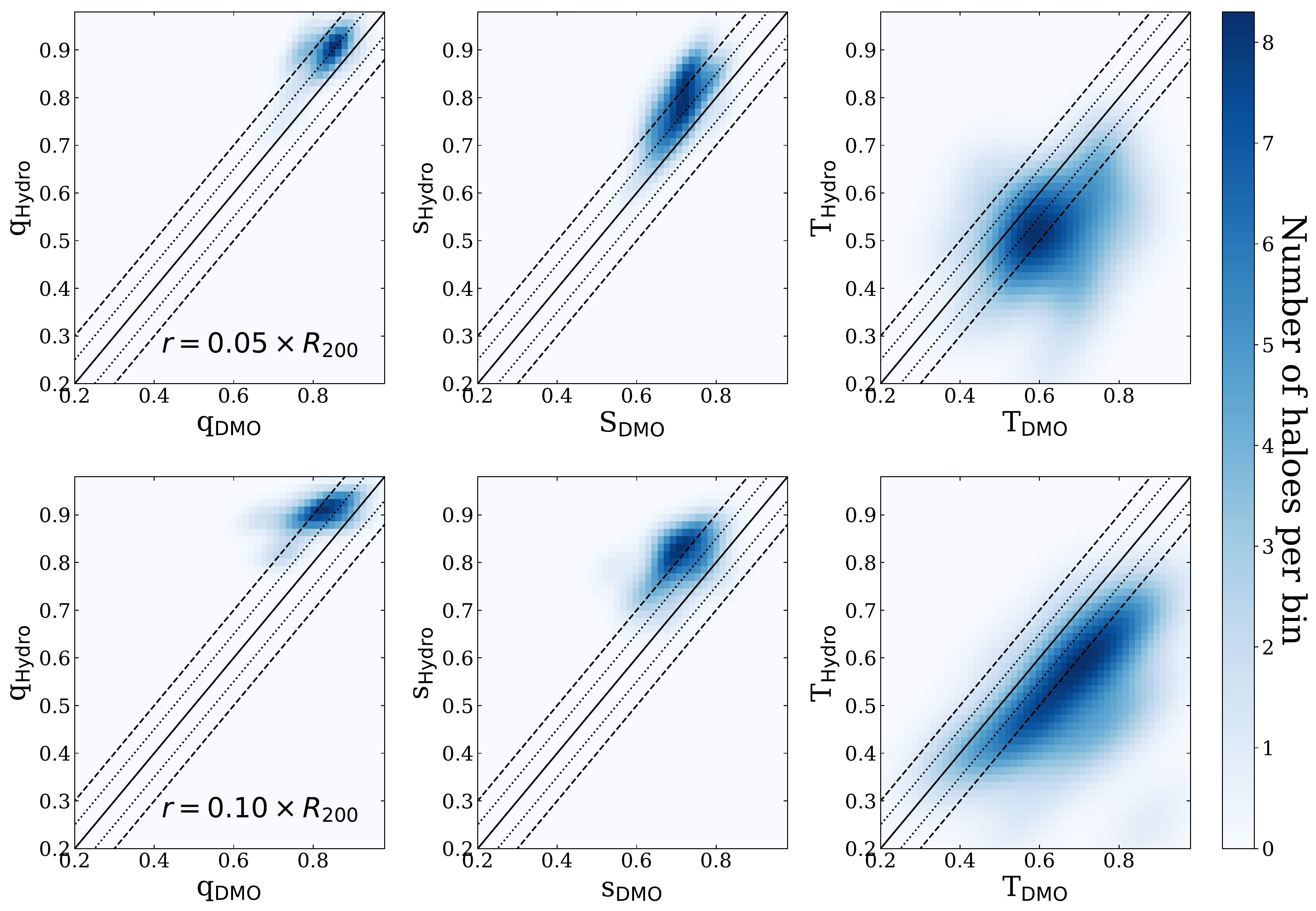}  
\includegraphics[width=\columnwidth]{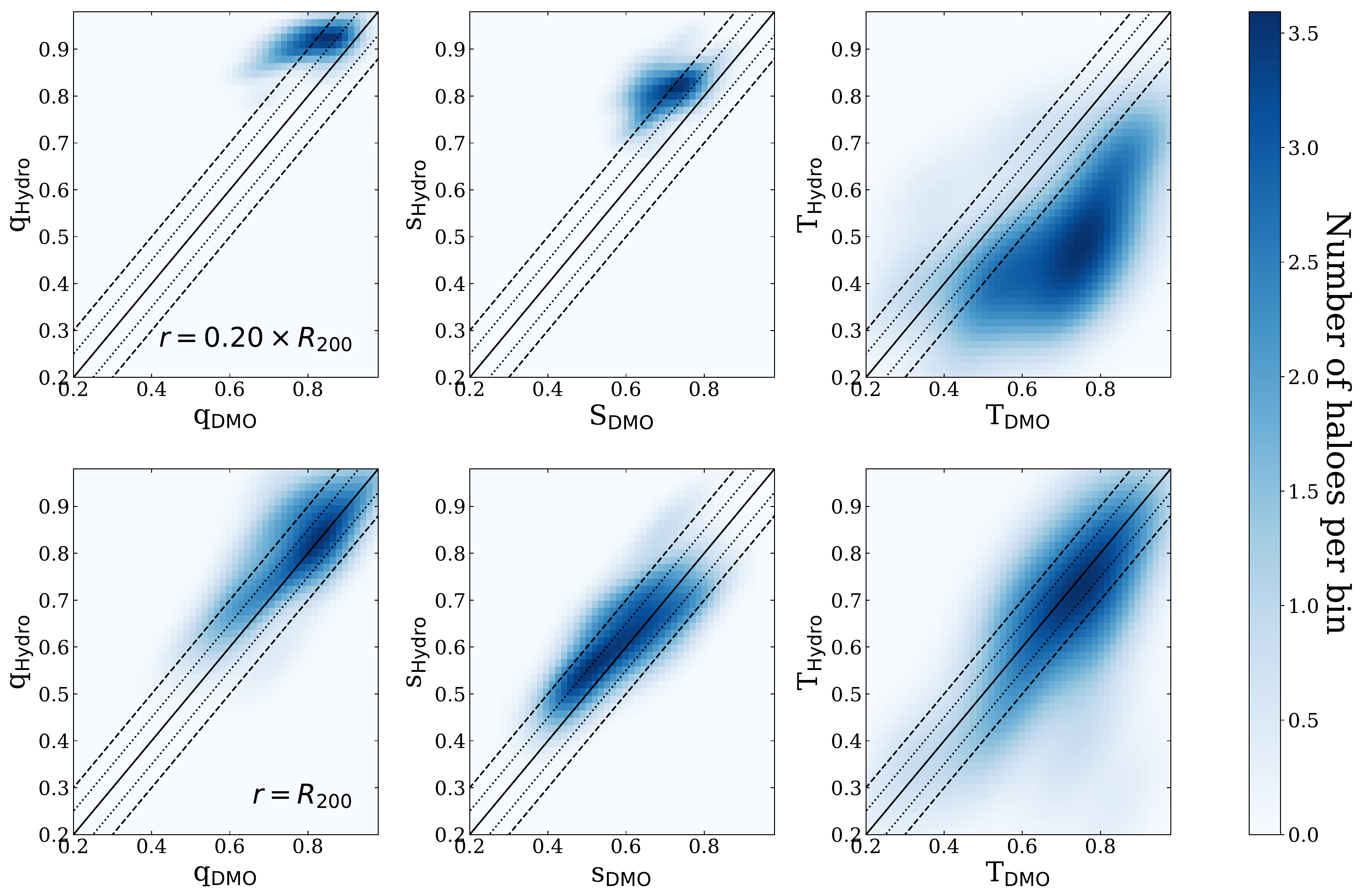}  
\caption{$\rm 2D$ histogram  of the DM halo shape parameters for Hydro versus DMo simulations measured at $\rm r=0.05R_{\rm 200}$ (top left), $\rm r=0.10R_{\rm 200}$ (bottom left), $\rm r=0.20R_{\rm 200}$ (top right) and $\rm r=R_{\rm 200}$ (bottom right) for the Fenix haloes. The 1:1 relations are shown (solid lines) together with the 5 and 10 percent deviations (dotted lines) for both Hydro and DMo runs. The inner haloes from the Hydro run are more spherical and oblate than the DMo counterparts. At the virial radius, the effect of baryons is less significant whereas the maximum deviations are also found for $\rm r=0.20R_{\rm 200}$.}
\label{fig:fig4}
\end{figure*}

We present in Fig.~\ref{fig:fig4} the $\rm 2D$ histograms comparing the shape parameters for Hydro and DMo haloes for the Fenix simulation at the three selected radii: $\rm r=0.05R_{\rm 200}$, $\rm r=0.10R_{\rm 200}$, $\rm r=0.20R_{\rm 200}$ and the virial radius $\rm R_{\rm 200}$. As well as \textsc{eagle} haloes, in the inner regions, the largest deviation of  Hydro haloes shape haloes from DMo ones is detected at $\rm r=0.20R_{\rm 200}$.

\begin{figure*}
  \centering
  \includegraphics[width=\columnwidth]{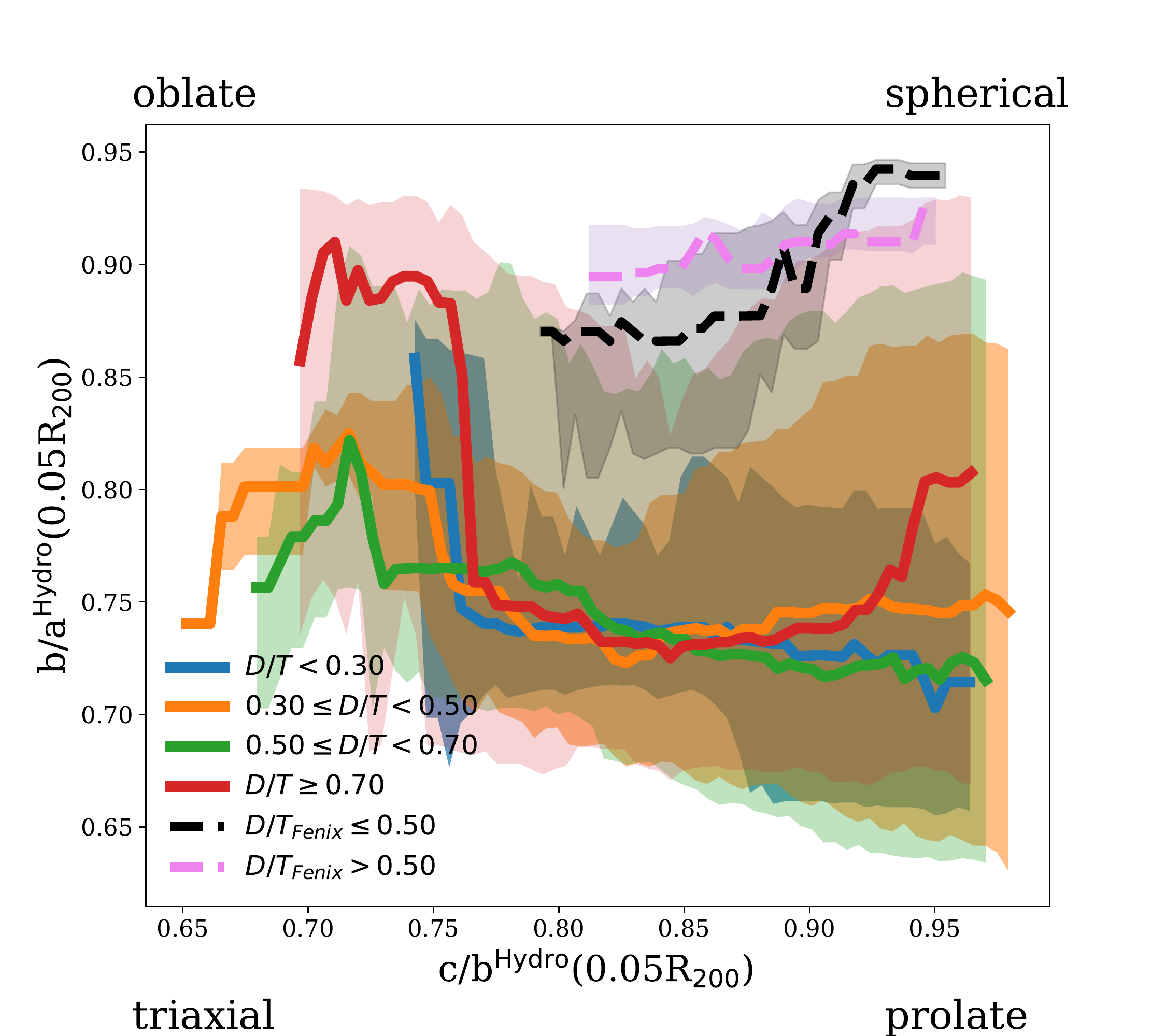}
  \includegraphics[width=\columnwidth]{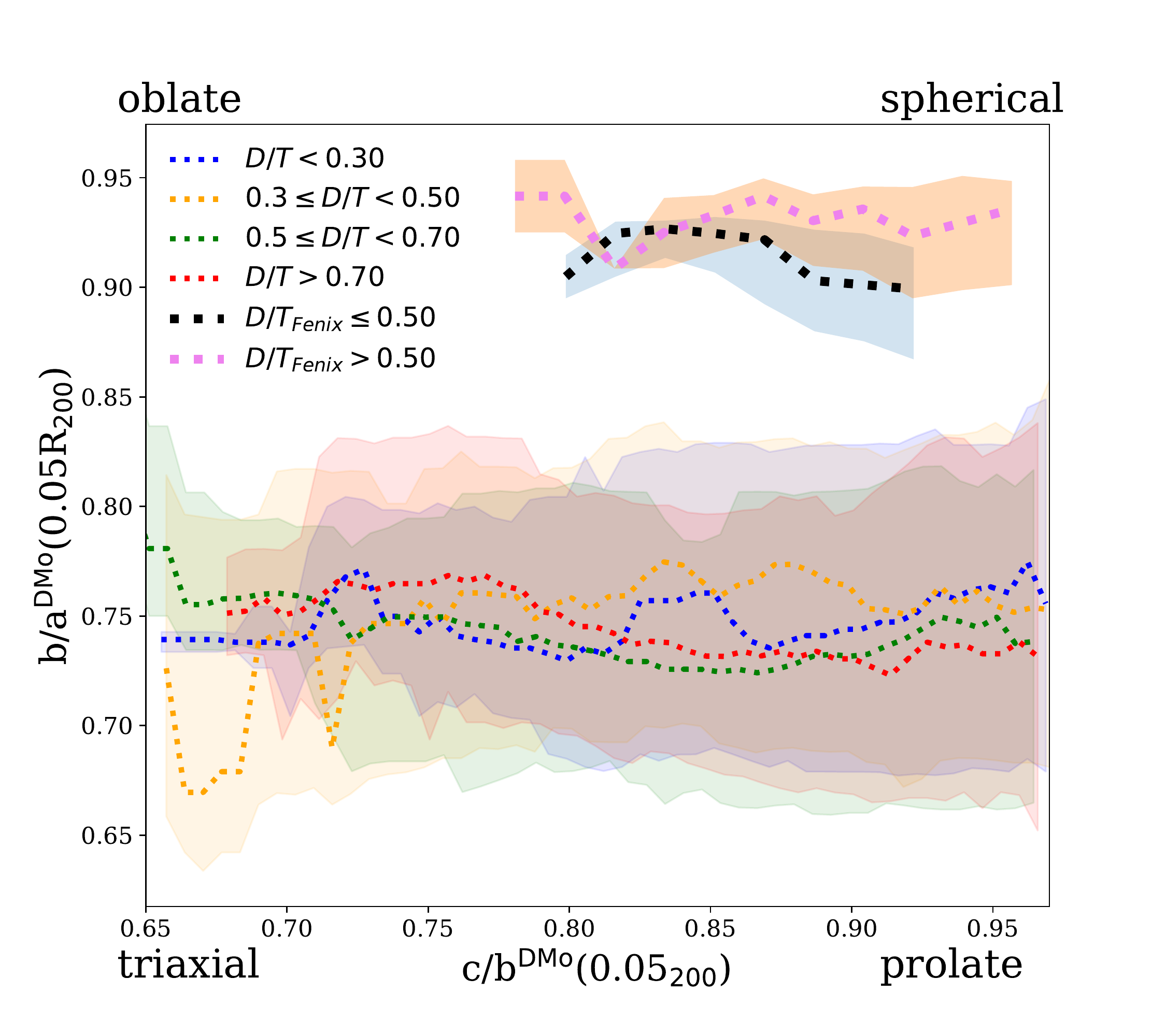}
\caption{The distribution of {\sc eagle} and Fenix haloes as a function of their inner halo axial ratios measured at 5 percent of the virial radius. Individual haloes are coloured by the $\rm D/T$ fraction. Solid lines indicate the median of the distribution for bines of $\rm D/T$ and in dotted lines the their counterpart in the DMo simulation. The shaded areas denote the $\rm 25th$ and $\rm 75th$ central quartiles.}
\label{fig:fig12}
\end{figure*}

\begin{figure*}
  \centering
  \includegraphics[width=\columnwidth]{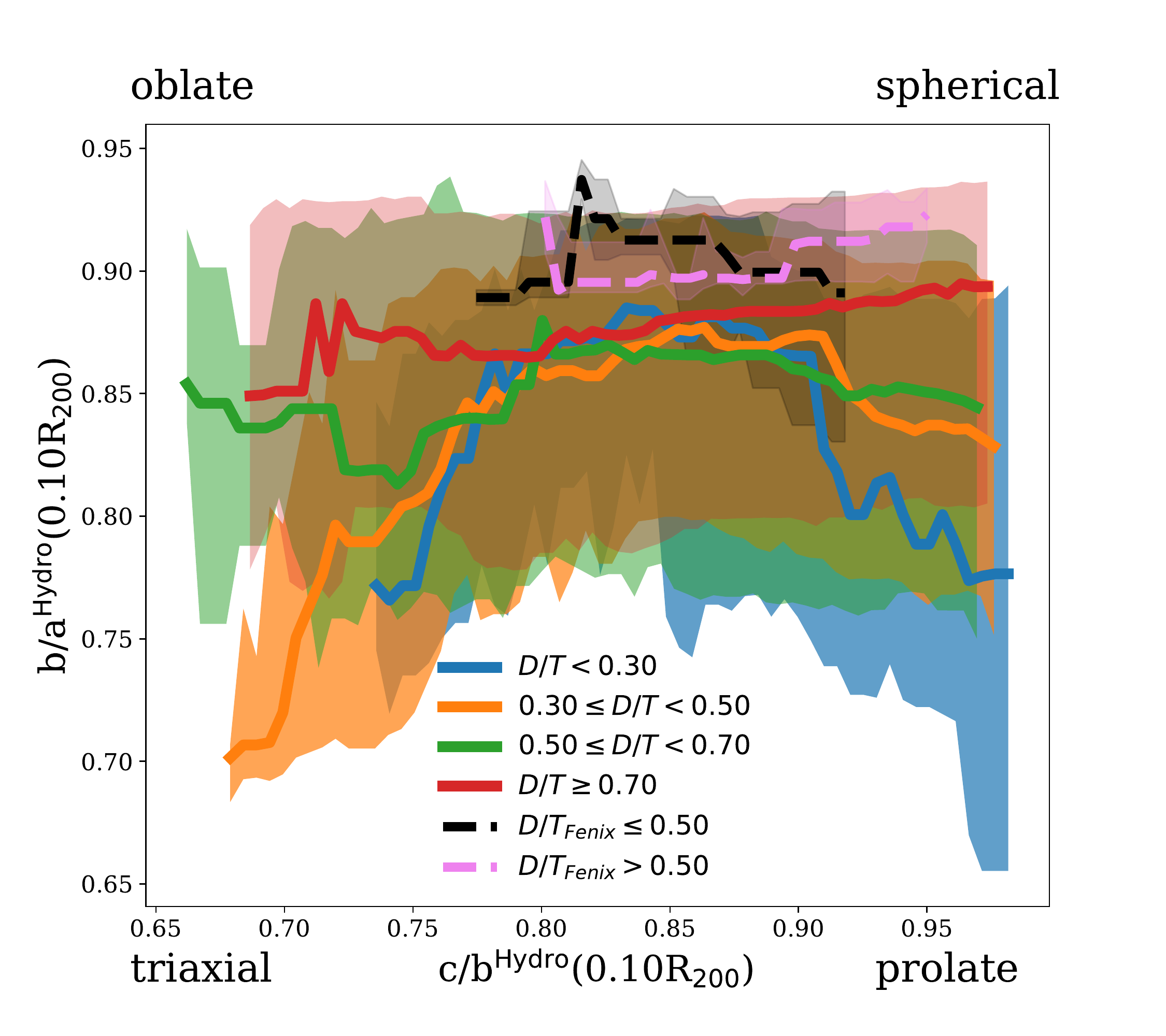}
  \includegraphics[width=\columnwidth]{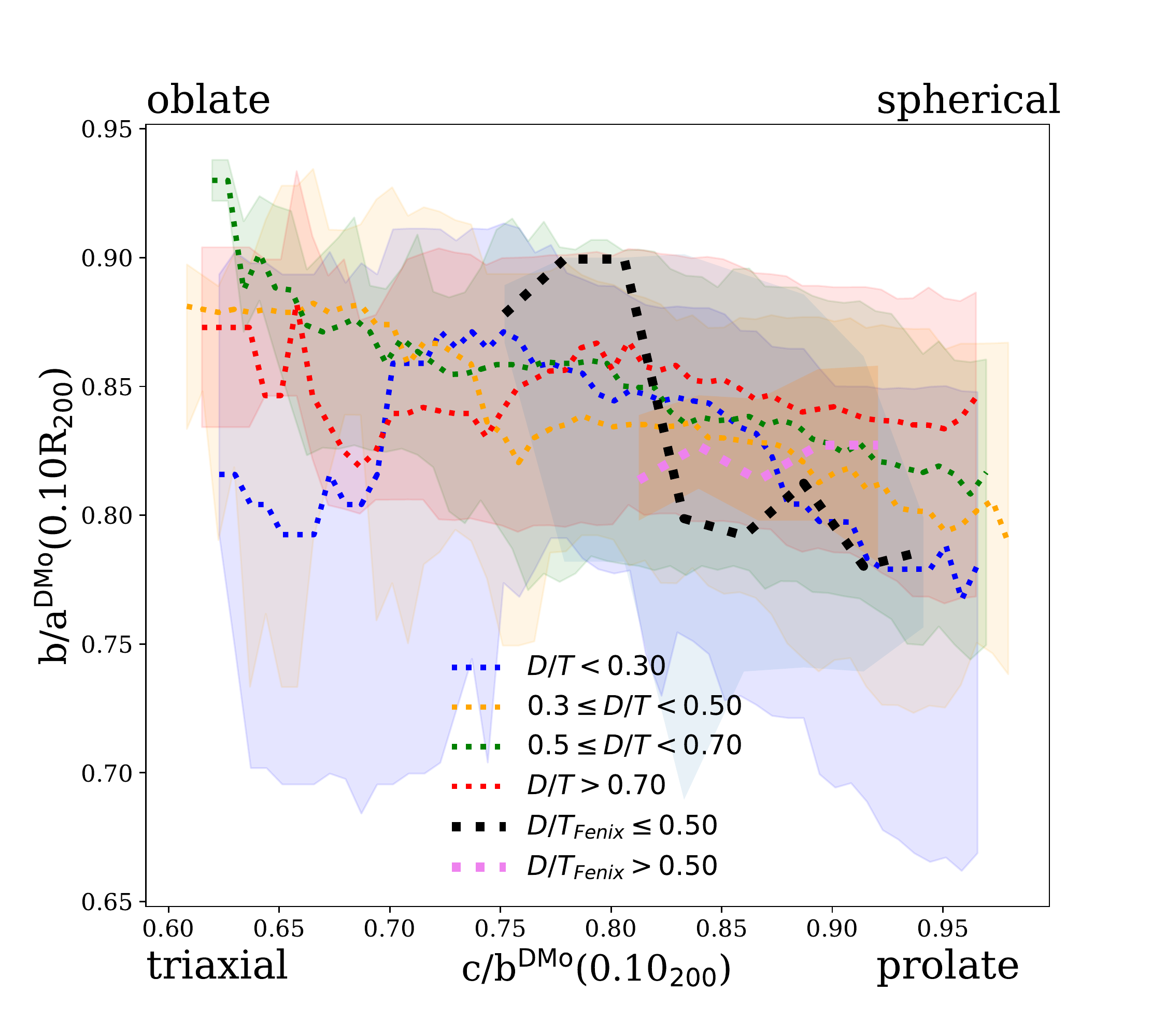}
\caption{Same as Fig. \ref{fig:fig12} at 10 percent of the virial radius.} 
\label{fig:fig13}
\end{figure*}

We also extend the analysis presented in Section~\ref{sec:shapemorphology}, by showing the distribution of the inner axial ratios for {\sc eagle} and Fenix haloes at $\mathrm{0.05R_{200}}$ (Fig.~\ref{fig:fig12}) and $\mathrm{0.10R_{200}}$ (Fig.~\ref{fig:fig13}).

\bsp	
\label{lastpage}
\end{document}